\documentclass[twocolumn]{aastex63}
\usepackage{graphics,graphicx}
\hypersetup{urlcolor=blue}
\usepackage{natbib}
\usepackage[outdir=./]{epstopdf}
\usepackage{textcomp,gensymb}
\usepackage{xfrac}
\usepackage{xcolor}

\newcommand{\name}{HD 63433}

\newcommand{\mps}{m\,s$^{-1}$}

\newcommand{\vsini}{$v\sin{i_*}$}

\newcommand{\logg}{$log~g$ }

\newcommand{\um}{$\mu$m}
\newcommand{\fbol}{$F_{\mathrm{bol}}$}

\newcommand{\teff}{\ensuremath{T_{\text{eff}}}}
\newcommand\kms{km~s$^{-1}$}

\newcommand{\tess}{\textit{TESS}}
\newcommand{\ktwo}{{\textit K2}}
\newcommand{\re}{$R_{\oplus}$}
\newcommand{\gaia}{{\textit Gaia}}

\accepted{August 10, 2020}
\submitjournal{AJ}

\shorttitle{Planets in the 400\,Myr Ursa Major Group}
\shortauthors{Mann et al.}
\graphicspath{{./}{}}

\begin{document}

\title{TESS Hunt for Young and Maturing Exoplanets (THYME) III: a two-planet system in the 400 Myr Ursa Major Group}

\correspondingauthor{Andrew W. Mann}
\email{awmann@unc.edu}

\author[0000-0003-3654-1602]{Andrew W. Mann}%
\affiliation{Department of Physics and Astronomy, The University of North Carolina at Chapel Hill, Chapel Hill, NC 27599, USA} 

\author[0000-0002-5099-8185]{Marshall C. Johnson}%
\affiliation{Las Cumbres Observatory, 6740 Cortona Dr., Ste. 102, Goleta, CA 93117, USA}

\author[0000-0001-7246-5438]{Andrew Vanderburg}%
\altaffiliation{NASA Sagan Fellow}
\affiliation{Department of Astronomy, The University of Texas at Austin, Austin, TX 78712, USA}

\author[0000-0001-9811-568X]{Adam L. Kraus}%
\affiliation{Department of Astronomy, The University of Texas at Austin, Austin, TX 78712, USA}

\author[0000-0001-9982-1332]{Aaron C. Rizzuto}%
\altaffiliation{51 Pegasi b Fellow}
\affiliation{Department of Astronomy, The University of Texas at Austin, Austin, TX 78712, USA}

\author[0000-0001-7336-7725]{Mackenna L. Wood}%
\affiliation{Department of Physics and Astronomy, The University of North Carolina at Chapel Hill, Chapel Hill, NC 27599, USA} 

\author[0000-0002-9446-9250]{Jonathan L. Bush}%
\affiliation{Department of Physics and Astronomy, The University of North Carolina at Chapel Hill, Chapel Hill, NC 27599, USA} 

\author[0000-0003-1337-723X]{Keighley Rockcliffe}%
\affiliation{Department of Physics and Astronomy, Dartmouth College, Hanover, NH 03755, USA}

\author[0000-0003-4150-841X]{Elisabeth R. Newton}%
\affiliation{Department of Physics and Astronomy, Dartmouth College, Hanover, NH 03755, USA}

\author[0000-0001-9911-7388]{David~W.~Latham}%
\affiliation{Center for Astrophysics ${\rm \mid}$ Harvard {\rm \&} Smithsonian, 60 Garden Street, Cambridge, MA 02138, USA}

\author[0000-0003-2008-1488]{Eric E. Mamajek}%
\affiliation{Jet Propulsion Laboratory, California Institute of Technology, 4800 Oak Grove Dr., Pasadena, CA 91109, USA} 

\author[0000-0002-4891-3517]{George Zhou}%
\altaffiliation{Hubble Fellow}
\affiliation{Center for Astrophysics ${\rm \mid}$ Harvard {\rm \&} Smithsonian, 60 Garden Street, Cambridge, MA 02138, USA}

\author[0000-0002-8964-8377]{Samuel N. Quinn}%
\affiliation{Center for Astrophysics ${\rm \mid}$ Harvard {\rm \&} Smithsonian, 60 Garden Street, Cambridge, MA 02138, USA}

\author[0000-0001-5729-6576]{Pa Chia Thao}%
\altaffiliation{NSF GRFP Fellow} 
\affiliation{Department of Physics and Astronomy, The University of North Carolina at Chapel Hill, Chapel Hill, NC 27599, USA} 

\author[0000-0002-4638-3495]{Serena Benatti} %
\affiliation{INAF -- Osservatorio Astronomico di Palermo, Piazza del Parlamento 1, I-90134, Palermo, Italy}

\author{Rosario Cosentino}%
\affiliation{Fundaci\'on Galileo Galilei-INAF, Rambla Jos\'e Ana Fernandez P\'erez 7, 38712 Bre\~na Baja, TF, Spain}

\author[0000-0001-8613-2589]{Silvano Desidera}%
\affiliation{INAF – Osservatorio Astronomico di Padova, Vicolo dell’Osservatorio 5, I-35122, Padova, Italy}

\author{Avet Harutyunyan}%
\affiliation{Fundaci\'on Galileo Galilei-INAF, Rambla Jos\'e Ana Fernandez P\'erez 7, 38712 Bre\~na Baja, TF, Spain}

\author{Christophe Lovis}%
\affiliation{Observatoire Astronomique de l'Universit\'e de Gen\`eve, 51 chemin des Maillettes, 1290 Versoix, Switzerland}

\author[0000-0001-7254-4363]{Annelies Mortier}%
\affiliation{Astrophysics Group, Cavendish Laboratory, University of Cambridge,
  J.J. Thomson Avenue, Cambridge CB3 0HE, UK}

\author{Francesco A. Pepe}%
\affiliation{Observatoire Astronomique de l'Universit\'e de Gen\`eve, 51 chemin des Maillettes, 1290 Versoix, Switzerland}
  
\author[0000-0003-1200-0473]{Ennio Poretti}%
\affiliation{Fundaci\'on Galileo Galilei-INAF, Rambla Jos\'e Ana Fernandez P\'erez 7, 38712 Bre\~na Baja, TF, Spain}
\affiliation{INAF-Osservatorio Astronomico di Brera, via E. Bianchi 46, 23807
Merate (LC), Italy}
  
\author[0000-0001-8749-1962]{Thomas G. Wilson} %
\affiliation{School of Physics and Astronomy, University of St Andrews, North Haugh, St Andrews, Fife, KY16 9SS, UK}

\author[0000-0002-2607-138X]{Martti H. Kristiansen}%
\affiliation{Brorfelde Observatory, Observator Gyldenkernes Vej 7, DK-4340 T\o{}ll\o{}se, Denmark}
\affiliation{DTU Space, National Space Institute, Technical University of Denmark, Elektrovej 327, DK-2800 Lyngby, Denmark}

\author[0000-0002-5665-1879]{Robert Gagliano}%
\affiliation{Amateur Astronomer, Glendale, Arizona, USA}

\author[0000-0003-3988-3245]{Thomas Jacobs}%
\affiliation{Amateur Astronomer, 12812 SE 69th Place, Bellevue WA, USA}

\author[0000-0002-8527-2114]{Daryll M. LaCourse}%
\affiliation{Amateur Astronomer, 7507 52nd Pl NE, Marysville, WA, 98270, USA}

\author{Mark Omohundro}%
\affiliation{Citizen Scientist, c/o Zooniverse, Department of Physics, University of Oxford, Denys Wilkinson Building, Keble Road, Oxford, OX1 3RH, UK}

\author{Hans Martin Schwengeler}%
\affiliation{Citizen Scientist, Planet Hunter, Bottmingen, Switzerland}

\author{Ivan A. Terentev}  
\affiliation{Citizen Scientist, Planet Hunter, Petrozavodsk, Russia}

\author[0000-0002-7084-0529]{Stephen R. Kane}%
\affiliation{Department of Earth and Planetary Sciences, University of California, Riverside, CA 92521, USA}

\author[0000-0002-0139-4756]{Michelle L. Hill}%
\affiliation{Department of Earth and Planetary Sciences, University of California, Riverside, CA 92521, USA} 


\author[0000-0003-2935-7196]{Markus Rabus}%
\affiliation{Las Cumbres Observatory, 6740 Cortona Dr., Ste. 102, Goleta, CA 93117, USA}
\affiliation{Department of Physics, University of California, Santa Barbara, CA 93106-9530, USA}

\author[0000-0002-9789-5474]{Gilbert A. Esquerdo} %
\affiliation{Center for Astrophysics ${\rm \mid}$ Harvard {\rm \&} Smithsonian, 60 Garden Street, Cambridge, MA 02138, USA}

\author{Perry Berlind}%
\affiliation{Center for Astrophysics ${\rm \mid}$ Harvard {\rm \&} Smithsonian, 60 Garden Street, Cambridge, MA 02138, USA}


\author[0000-0001-6588-9574]{Karen A.\ Collins}%
\affiliation{Center for Astrophysics ${\rm \mid}$ Harvard {\rm \&} Smithsonian, 60 Garden Street, Cambridge, MA 02138, USA}

\author[0000-0001-7809-1457]{Gabriel Murawski}%
\affiliation{SOTES Private Observatory, Suwalki, Poland}
\affiliation{WWVSH (World Wide Variable Star Hunters), Abu Dhabi, United Arab Emirates}

\author[0000-0003-0614-2571]{Nezar Hazam Sallam} %
\affiliation{WWVSH (World Wide Variable Star Hunters), Abu Dhabi, United Arab Emirates}

\author[0000-0002-0169-0766]{Michael M. Aitken}%
\affiliation{WWVSH (World Wide Variable Star Hunters), Abu Dhabi, United Arab Emirates}

\author[0000-0001-8879-7138]{Bob Massey}%
\affiliation{Villa `39 Observatory, Landers, CA 92285, USA}

\author[0000-0003-2058-6662]{George~R.~Ricker}%
\affiliation{Department of Physics and Kavli Institute for Astrophysics and Space Research, Massachusetts Institute of Technology, Cambridge, MA 02139, USA}

\author[0000-0001-6763-6562]{Roland~Vanderspek}%
\affiliation{Department of Physics and Kavli Institute for Astrophysics and Space Research, Massachusetts Institute of Technology, Cambridge, MA 02139, USA}

\author[0000-0002-6892-6948]{Sara~Seager}%
\affiliation{Department of Physics and Kavli Institute for Astrophysics and Space Research, Massachusetts Institute of Technology, Cambridge, MA 02139, USA}
\affiliation{Department of Earth, Atmospheric and Planetary Sciences, Massachusetts Institute of Technology, Cambridge, MA 02139, USA}
\affiliation{Department of Aeronautics and Astronautics, MIT, 77 Massachusetts Avenue, Cambridge, MA 02139, USA}

\author[0000-0002-4265-047X]{Joshua~N.~Winn}%
\affiliation{Department of Astrophysical Sciences, Princeton University, 4 Ivy Lane, Princeton, NJ 08544, USA}

\author[0000-0002-4715-9460]{Jon M. Jenkins}%
\affiliation{NASA Ames Research Center, Moffett Field, CA, 94035, USA}

\author[0000-0001-7139-2724]{Thomas~Barclay}%
\affiliation{NASA Goddard Space Flight Center, 8800 Greenbelt Road, Greenbelt, MD 20771, USA}
\affiliation{University of Maryland, Baltimore County, 1000 Hilltop Circle, Baltimore, MD 21250, USA}

\author[0000-0003-1963-9616]{Douglas A. Caldwell}%
\affiliation{SETI Institute, Mountain View, CA, 94043, USA}
\affiliation{NASA Ames Research Center, Moffett Field, CA, 94035, USA}

\author[0000-0003-2313-467X]{Diana~Dragomir}%
\affiliation{Department of Physics and Astronomy, University of New Mexico, 1919 Lomas Blvd NE, Albuquerque, NM 87131, USA}

\author{John~P.~Doty}%
\affiliation{Noqsi Aerospace Ltd., 15 Blanchard Avenue, Billerica, MA 01821, USA}

\author[0000-0002-5322-2315]{Ana~Glidden}%
\affiliation{Department of Earth, Atmospheric and Planetary Sciences, Massachusetts Institute of Technology, Cambridge, MA 02139, USA}
\affiliation{Department of Physics and Kavli Institute for Astrophysics and Space Research, Massachusetts Institute of Technology, Cambridge, MA 02139, USA}

\author[0000-0002-1949-4720]{Peter Tenenbaum}%
\affiliation{SETI Institute, Mountain View, CA, 94043, USA}
\affiliation{NASA Ames Research Center, Moffett Field, CA, 94035, USA}

\author[0000-0002-5286-0251]{Guillermo~Torres}%
\affiliation{Center for Astrophysics ${\rm \mid}$ Harvard {\rm \&} Smithsonian, 60 Garden Street, Cambridge, MA 02138, USA}

\author[0000-0002-6778-7552]{Joseph D. Twicken}%
\affiliation{SETI Institute, Mountain View, CA, 94043, USA}
\affiliation{NASA Ames Research Center, Moffett Field, CA, 94035, USA}

\author[0000-0001-6213-8804]{Steven~Villanueva~Jr}%
\altaffiliation{Pappalardo Fellow}
\affiliation{Department of Physics and Kavli Institute for Astrophysics and Space Research, Massachusetts Institute of Technology, Cambridge, MA 02139, USA}

\begin{abstract}
Exoplanets can evolve significantly between birth and maturity, as their atmospheres, orbits, and structures are shaped by their environment. Young planets ($<$1 Gyr) offer an opportunity to probe the critical early stages of this evolution, where planets evolve the fastest. However, most of the known young planets orbit prohibitively faint stars. We present the discovery of two planets transiting HD 63433 (TOI 1726, TIC 130181866), a young Sun-like ($M_*=0.99\pm0.03$) star. Through kinematics, lithium abundance, and rotation, we confirm that HD 63433 is a member of the Ursa Major moving group ($\tau=414\pm23$\,Myr). Based on the TESS light curve and updated stellar parameters, we estimate the planet radii are $2.15\pm0.10R_\oplus$ and $2.67\pm0.12R_\oplus$, the orbital periods are 7.11 and 20.55\,days, and the orbital eccentricities are lower than about 0.2. Using HARPS-N velocities, we measure the Rossiter-McLaughlin signal of the inner planet, demonstrating the orbit is prograde.  Since the host star is bright (V=6.9), both planets are amenable to transmission spectroscopy, radial velocity measurements of their masses, and more precise determination of the stellar obliquity. This system is therefore poised to play an important role in our understanding of planetary system evolution in the first billion years after formation.

\end{abstract}


\section{Introduction} \label{sec:intro}
Over their lifetimes, the dynamical, structural, and atmospheric properties of planets are modified by their environment \citep[e.g.,][]{Kaib:2013, Ehrenreich2015} and internal processes \citep{Fortney2011, GinzburgCorepowered2018}. The simplest observational path to explore these processes is to compare the statistical properties of planets at different ages. Since the evolution is the most rapid in the first few hundred million years, planets with known ages $<$1\,Gyr are especially useful. 

With this in mind, the Zodiacal Exoplanets in Time Survey \citep[ZEIT; ][]{Mann2016a}, and its successor, the \tess\ Hunt for Young and Maturing Exoplanets \citep[THYME; ][]{Newton2019} set out to identify transiting planets in young clusters, moving groups, and star-forming regions with ages of 5-700\,Myr using light curves from the {\it K2} and {\it TESS} missions. Discoveries from these and similar surveys have found planets in diverse environments, from the 10-20\,Myr Sco-Cen OB association \citep{Rizzuto2020}, to the 45\,Myr Tucana-Horologium moving group \citep{Newton2019, Benatti_dstuc}, and as old as the 700\,Myr Hyades cluster \citep{Vanderburg2018}. More importantly, these discoveries have demonstrated that young planets are systematically larger than older planets of the same mass \citep{Obermeier2016, Mann:2018} and that at least some short-period planets migrate within the first 10\,Myr or form in situ \citep{Mann2016b, David2016b}. 

Studies of individual young systems can also be powerful, providing new insight into topics such as haze and cloud formation in young systems \citep[e.g.,][]{Gao_hazes, Thao2020}, photoevaporation and atmospheric escape \citep[e.g.,][Rockcliffe et al. in prep]{Gaidos:2020}, and exoplanet migration \citep[e.g.,][]{Mann2016b,David2016b}. In particular, measurement of spin-orbit misalignments via the Rossiter-McLaughlin (RM) effect are important for young and multiplanet systems to inform our understanding of their dynamical histories. \tess\ has already enabled the discovery of young planets around bright stars \citep{Newton2019}, facilitating spin-orbit alignment measurements \citep{Zhou20, Montet20}.

Citizen scientists have long played an important role in the discovery of important planetary systems, particularly from {\it Kepler}, {\it K2}, and {\it TESS} mission data. This is in large part due to the success of programs like Planet Hunters \citep{2013ApJ...768..127S, 2013ApJ...776...10W} and Exoplanet Explorers \citep{2018AJ....155...57C}. Young planets are no exception; citizen scientists aided in the discovery and characterizing of both K2-25b \citep[$\simeq$700\,Myr,][]{Mann2016a} and K2-233 \citep[$\sim$400\,Myr]{David:2018aa}.

Despite these recent exoplanet discoveries, the sample of transiting planets with known, young ages is still small ($\simeq30$ planets), and most of them orbit stars too faint for follow-up with existing precision radial velocity instruments (PRV). The sample is also heavily biased toward the extreme age ends of the survey, with most of the known planets at 700\,Myr or $<$50\,Myr \citep{Mann2017a, 2019ApJ...885L..12D}. 

We report the discovery of two young transiting planets, both with radii between 2$R_\oplus$ and 3$R_\oplus$. The host star (\name) is a bright ($V\simeq6.9$) member of the $\simeq400$\,Myr Ursa Major Moving group. \name\ is the third-brightest star (by optical magnitude) discovered to host a transiting planet using \tess\ data; the only brighter stars so far are $\pi$ Men \citep{Huang2018_pimen} and HR 858 \citep{Vanderburg2019}.

In Section \ref{sec:obs}, we present the discovery data from \tess, as well as follow-up and archival photometry and spectroscopy used to characterize the planets and stellar host. In Section \ref{sec:star}, we demonstrate that \name\ is a member of Ursa Major and update the stellar parameters (radius, mass, \teff, age, and rotation). We fit the \tess\ transit data and Rossiter-McLaughlin (RM) velocities to provide parameters of both planets, which we discuss in Section~\ref{sec:transit}. We detail our validation of the signals as planetary in origin in Section~\ref{sec:fpp} and discuss its dynamical stability in Section~\ref{sec:dyn}. We conclude in Section~\ref{sec:discussion} with a brief summary and discussion of future follow-up of \name bc and of the Ursa Major cluster more generally. 
 
\section{Observations}\label{sec:obs}

\subsection{TESS Photometry}\label{sec:tess}

The {\it TESS} mission \citep{2014SPIE.9143E..20R} observed TIC 130181866 (TOI 1726, HD 63433, HIP 38228) between 2019 December 24 and 2020 January 21 (Sector 20) using Camera 1. The target was proposed by three guest investigator programs (G022032, T. Metcalfe; G022038, R. Roettenbacher; G022203, J. Ge) and hence has has 2-minute cadence data. The abstracts for these GI programs suggest \name\ was targeted because it is bright ($V\simeq7$ mag) and/or active. 

For our analysis, we used the Presearch Data Conditioning simple aperture photometry \citep[PDCSAP; ][]{Smith2012, Stumpe2014} \tess\ light curve produced by the Science Process Operations centre \citep[SPOC; ][]{Jenkins:2016} and available through the Mikulski Archive for Space Telescopes (MAST)\footnote{\url{https://mast.stsci.edu/portal/Mashup/Clients/Mast/Portal.html}}. We only included data points with \texttt{DQUALITY=0}, i.e., those with no flags from the SPOC pipeline. No obvious flares were present in the data, so we did no further data processing. 

\subsection{Ground-Based Photometry}\label{sec:phot}

We obtained time series photometry during three predicted transits of \name~b using ground-based facilities in order to rule out nearby eclipsing binaries (NEBs) which could be the source of the transit signal. We observed an egress on 2020 February 22 UT, and a full transit on 2020 February 29, both in $r$ with the 0.6 m World Wide Variable Star Hunters (WWVSH) telescope in Abu Dhabi, United Arab Emirates. Both transits were observed under clear (near-photometric) conditions. The telescope is equipped with a Finger Lakes Instrumentation FLI 16803 camera, giving a pixel scale of 0.47'' pixel$^{-1}$. We obtained 118 and 350 exposures with an exposure length of 180 and 90 seconds for the two observations, respectively. We also observed a full transit of the 2020 February 29 event with one of the 1\,m telescopes at the Las Cumbres Observatory \citep[LCOGT;][]{Brown13} node at the South African Astronomical Observatory, South Africa under clear conditions. We observed in the $z_s$ band using a Sinistro camera, giving a pixel scale of 0.389'' pixel$^{-1}$. We obtained 161 60-second exposures, which were reduced using the BANZAI pipeline \citep{McCully18}. In all cases, we deliberately saturated the target star in order to search for faint NEBs.

We performed aperture photometry on all three datasets using the \texttt{AstroImageJ} package \citep[AIJ;][]{Collins17}. TIC 130181879 (\tess\ magnitude 13.1) and TIC 130181877 (\tess\ magnitude 14.6) are the only two stars within 2.5\arcmin\ of \name\ that are bright enough to cause the \tess\ detection, so we extracted light curves for both. We used the aperture size and set of comparison stars that yielded the best precision for each observation and star of interest. For both WWVSH observations, we used a 7-pixel (3.3\arcsec) radius circular aperture to extract the source and an annulus with a 12-pixel (5.6\arcsec) inner radius and a 17-pixel (8\arcsec) outer radius for the sky. For the LCO photometry, we used a 6-pixel (2.3\arcsec) radius circular aperture for the source and a annulus with a 16-pixel (6.2\arcsec) inner radius and a 23-pixel (9\arcsec) outer radius for the sky. For all observations, we centered the apertures on the source, weighted all pixels within the aperture equally, and used the same aperture setup for TIC 130181879 and TIC 130181877. The extracted light curves (including for additional nearby faint sources), field overlays, and further information on this follow-up can be found on on ExoFOP-TESS\footnote{\url{https://exofop.ipac.caltech.edu/tess/target.php?id=130181866}}.

To reproduce the observed transit depths, eclipses around either nearby star would need to be large ($\simeq$20\%), but no binary was detected down to $\lesssim1\%$. Thus, the observations ruled out any NEB scenario consistent with the observed transit.

\subsection{Spectroscopy}

We utilized new and archival high-resolution spectra and radial velocity measurements of \name\ in our analysis. We list all of the radial velocity data in Tables~\ref{tab:RVs} and \ref{tab:RM}.

\begin{deluxetable}{lccr}
\centering
\tabletypesize{\scriptsize}
\tablewidth{0pt}
\tablecaption{Radial Velocity Measurements of \name \label{tab:RVs}}
\tablehead{\colhead{BJD} & \colhead{$v$ (\kms)\tablenotemark{a}} & \colhead{$\sigma_v$ (\kms)\tablenotemark{b}} & \colhead{Instrument} }
\startdata
2450510.3603 & -15.798 & 0.023 & ELODIE \\ 
2450511.4079 & -15.831 & 0.023 & ELODIE \\ 
2451984.308 & -15.851 & 0.023 & ELODIE \\ 
2456945.67006 & -15.811 & 0.002 & SOPHIE \\ 
2457102.36074 & -15.817 & 0.002 & SOPHIE \\ 
2457099.37892 & -15.757 & 0.002 & SOPHIE \\ 
2457059.53146 & -15.847 & 0.003 & SOPHIE \\ 
2457492.3052 & -15.841 & 0.002 & SOPHIE \\ 
2457490.30522 & -15.858 & 0.002 & SOPHIE \\ 
2457448.41586 & -15.854 & 0.002 & SOPHIE \\ 
2457444.45015 & -15.774 & 0.002 & SOPHIE \\ 
2456386.31072 & -15.846 & 0.001 & SOPHIE \\ 
2456388.34133 & -15.774 & 0.001 & SOPHIE \\ 
2456390.3025 & -15.871 & 0.002 & SOPHIE \\ 
2456383.32005 & -15.863 & 0.003 & SOPHIE \\ 
2456388.29898 & -15.777 & 0.002 & SOPHIE \\ 
\hline 
2450831.81836 & \phantom{-0}0.021 & 0.008 & Hamilton \\ 
2450854.79102 & \phantom{0}-0.009 & 0.009 & Hamilton \\ 
2451469.00303 & \phantom{0}-0.007 & 0.005 & Hamilton \\ 
2453014.84277 & \phantom{-0}0.040 & 0.006 & Hamilton\\ 
2453033.81543 & \phantom{0}-0.006 & 0.005 & Hamilton \\ 
2453068.72168 & \phantom{-0}0.011 & 0.006 & Hamilton \\ 
2453388.79327 & \phantom{0}-0.012 & 0.005 & Hamilton \\ 
2453390.88716 & \phantom{-0}0.003 & 0.005 & Hamilton \\ 
2454783.95126 & \phantom{0}-0.046 & 0.006 & Hamilton \\ 
2454865.81449 & \phantom{-0}0.031 & 0.005 & Hamilton \\ 
2455846.99256 & \phantom{0}-0.016 & 0.006 & Hamilton \\ 
\hline 
2458900.693564 & -15.838 & 0.028 & TRES \\
2458903.822341 & -15.867 & 0.028 & TRES \\
\hline 
2458906.27953 & -15.740 & \ldots & NRES \\ 
2458908.21348 & -16.060 & \ldots & NRES \\ 
2458912.24069 & -16.110 & \ldots & NRES \\ 
\enddata
\tablenotetext{a}{The Hamilton (Lick) radial velocities are relative, whereas the other radial velocities are on an absolute frame (although instrumental offsets may still be present).}
\tablenotetext{b}{RV errors are likely underestimated due to missing terms (e.g., from stellar jitter). The NRES pipeline does not currently estimate radial velocity} uncertainties. 
\end{deluxetable}

\begin{deluxetable}{lccr}
\centering
\tabletypesize{\scriptsize}
\tablewidth{0pt}
\tablecaption{HARPS-N Rossiter-McLaughlin Velocity Measurements\label{tab:RM}}
\tablehead{\colhead{BJD} & \colhead{$v$ (\kms)} & \colhead{$\sigma_v$ (\kms)} }
\startdata
2458916.36135 & -15.7481 & 0.0017  \\ 
2458916.36645 & -15.7495 & 0.0014   \\ 
2458916.37178 & -15.7477 & 0.0013  \\ 
2458916.37670 & -15.7467 & 0.0014   \\ 
2458916.38178 & -15.7478 & 0.0013  \\ 
2458916.38684 & -15.7460 & 0.0015   \\ 
2458916.39180 & -15.7444 & 0.0016   \\ 
2458916.39972 & -15.7451 & 0.0019  \\ 
2458916.41321 & -15.7448 & 0.0012  \\ 
2458916.41813 & -15.7415 & 0.0013   \\ 
2458916.42337 & -15.7455 & 0.0016  \\ 
2458916.42875 & -15.7434 & 0.0017  \\ 
2458916.43361 & -15.7439 & 0.0016  \\ 
2458916.43867 & -15.7454 & 0.0016  \\ 
2458916.44392 & -15.7447 & 0.0016   \\ 
2458916.44898 & -15.7500 & 0.0017 \\ 
2458916.45424 & -15.7472 & 0.0019  \\ 
2458916.45942 & -15.7453 & 0.0018  \\ 
2458916.46442 & -15.7481 & 0.0019  \\ 
2458916.46953 & -15.7492 & 0.0020  \\ 
2458916.47516 & -15.7471 & 0.0017  \\ 
2458916.48009 & -15.7482 & 0.0016  \\ 
2458916.48524 & -15.7474 & 0.0019  \\ 
2458916.49000 & -15.7485 & 0.0022   \\ 
2458916.49554 & -15.7475 & 0.0020  \\ 
2458916.50089 & -15.7479 & 0.0020  \\ 
2458916.50599 & -15.7490 & 0.0021   \\ 
2458916.51085 & -15.7474 & 0.0015   \\ 
2458916.51611 & -15.7481 & 0.0018  \\ 
2458916.52145 & -15.7463 & 0.0020  \\ 
2458916.52641 & -15.7463 & 0.0017  \\ 
2458916.53135 & -15.7450 & 0.0020   \\ 
2458916.53688 & -15.7433 & 0.0017  \\ 
2458916.54188 & -15.7458 & 0.0014  \\ 
2458916.54691 & -15.7432 & 0.0014  \\ 
2458916.55229 & -15.7480 & 0.0015   \\ 
2458916.55738 & -15.7481 & 0.0015  \\ 
2458916.56268 & -15.7470 & 0.0014   \\ 
2458916.56791 & -15.7451 & 0.0015  \\ 
2458916.57300 & -15.7463 & 0.0014 \\ 
2458916.57768 & -15.7467 & 0.0014 \\ 
2458916.58336 & -15.7472 & 0.0016 \\ 
2458916.58844 & -15.7470 & 0.0014  \\ 
\enddata
\end{deluxetable}

\subsubsection{LCOGT/NRES}

We obtained three spectra of \name\ using the LCOGT Network of Robotic Echelle Spectrographs \citep[NRES;][]{Siverd18} between 2020 Feb. 26 and Mar. 3 UT. All observations were taken under thin cloud-cover or clear conditions, using an exposure time of 900 seconds with the NRES unit at the Wise Observatory, Israel. NRES is a set of four identical cross-dispersed echelle spectrographs which are fiber-fed by 1\,m telescopes in the LCOGT network. NRES provides a resolving power of $R=53,000$ over the range $3800-8600$ \AA. The spectra were reduced, extracted, and wavelength calibrated using the standard NRES pipeline\footnote{\url{https://lco.global/documentation/data/nres-pipeline/}}. We measured radial velocities from the spectra using cross-correlation within the NRES Stage2 pipeline, and measured stellar parameters from the SpecMatch-Synth code\footnote{\url{https://github.com/petigura/specmatch-syn}}. The NRES spectra show no significant radial velocity shift between epochs, and no evidence of double lines or other indications of a false positive. 

\subsubsection{Tillinghast/TRES}

We obtained two spectra of \name\ with the 1.5m Tillinghast Reflector and the Tillinghast Reflector Echelle Spectrograph \citep[TRES;][]{Furesz08} located at Fred Lawrence Whipple Observatory, Arizona, USA. TRES is a cross-dispersed echelle spectrograph and delivers a resolving power of $R=44,000$ over the range $3900-9100$ \AA. We obtained one spectrum each on 2020 February 21 and 24 UT, near opposite quadratures of the orbit of \name~b. 

We reduced the TRES data and derived radial velocities using the standard TRES pipeline as described in \citep{2010ApJ...720.1118B}. We cross-correlated the extracted spectrum against a rotating synthetic spectrum with parameters similar to \name. For this, we used all 21 orders of the TRES spectrum, spanning 4140--6280\AA, only avoiding regions of high telluric contamination and variable features (e.g. Balmer lines). We then determined the instrumental noise floor from nightly observations of bright radial velocity standard stars and added this uncertainty in quadrature with the internal error estimates (derived from the variation between orders). To ensure our velocities are on the absolute scale from \citet{Nidever:2002}, we used the values derived from the Mg~b order (5190\AA) and applied an offset to account for the difference between standard star radial velocities determined in an identical manner and the velocities reported by \citet{Nidever:2002} for the same standards. The uncertainty in shifting to this absolute scale (and of the scale itself) is of order 0.1 km/s; errors on relative velocities are 0.028 km/s. Like the NRES data, the TRES spectra show no evidence of large velocity shifts or multiple lines that could indicate a binary.

\subsubsection{Goodman/SOAR}\label{sec:goodman}

To aid our spectral energy distribution fits (Section~\ref{sec:SED}), we obtained spectra of \name\ with the Goodman High-Throughput Spectrograph \citep{Goodman} on the Southern Astrophysical Research (SOAR) 4.1 m telescope located at Cerro Pachón, Chile. On 2020 March 6 (UT) and under photometric conditions, we took five spectra of \name, each with an exposure time of 5s using the red camera, the 1200 l/mm grating in the M5 setup, and the 0.46\arcsec\ slit rotated to the parallactic angle. This setup yielded a resolving power of $R \simeq 5900$ spanning 6250--7500\AA. For calibration, we obtained Ne arc lamps taken throughout the night (to account for drifts in the wavelength solution), as well as standard calibration data (dome/quartz flats and biases) taken during the afternoon. 

We performed bias subtraction, flat fielding, optimal extraction of the target spectrum, and found the wavelength solution using a 4th-order polynomial derived from the Ne lamp data. We then stacked the five extracted spectra using the robust weighted mean (for outlier removal). The stacked spectrum had a signal-to-noise ratio of $200-300$ over the full wavelength range. 

\subsubsection{HARPS-N/TNG}

With the aim of detecting the Rossiter-McLaughlin (RM) effect, we observed \name\ during the predicted transit of planet b on the night of 2020 March 7/8 (UT), under photometric conditions, with the High Accuracy Radial velocity Planet Searcher for the Northern hemisphere \citep[HARPS-N; ][]{Cosentino:2012, Cosentino:2014} spectrograph installed at the Telescopio Nazionale Galileo (TNG) at the Roque de los Muchachos Observatory on La Palma, Canary Islands, Spain. HARPS-N is a high-resolution (R$\simeq$120,000) spectrograph encased in a vacuum vessel that controls temperature and pressure at levels required for $<$1~\mps\ instrumental drifts. To cover both the transit and at least 1h of out-of-transit baseline, we took 43 spectra, spanning 5.4h in total and each with a fixed exposure time of 420s. 

Radial velocities were extracted from the HARPS-N spectra with the standard pipeline that uses a weighted cross-correlation with the numerical mask matching the spectral type (G2) of the target \citep{Pepe2002}. Typical radial velocity uncertainties were between 1--3~\mps. 

\subsection{Archival Velocities}\label{sec:rv}
Between 1997 March and 2016 April, \name\ was observed 15 times from the 1.93m telescope at the Haute-Provence Observatory located in France. The first three were taken with the ELODIE high-resolution spectrograph \citep{ELODIE} and the next 12 were taken by ELODIE's replacement, SOPHIE \citep{Perruchot:2008}. We retrieved the spectra and barycentric radial velocities given on the SOPHIE/ELODIE archives \citep{ELODIEarchive}\footnote{\url{http://atlas.obs-hp.fr/sophie/} \&  \url{http://atlas.obs-hp.fr/elodie/}}. To correct for differences in the zero-point between ELODIE and SOPHIE, we apply an offset of 87$\pm$23\,\mps\ to ELODIE velocities as described in \citet{2012A&A...545A..55B}. SOPHIE velocities were all taken after the upgrade to SOPHIE+ \citep{2013A&A...549A..49B} and have formal uncertainties of 1-3\mps, not including stellar jitter or long-term drift in the instrument \citep[$\simeq$5\mps; ][]{2015A&A...581A..38C}. 

As part of the Lick planet search program \name\ was observed 11 times between 1998 January and 2011 December using the Hamilton Spectrograph and iodine cell \citep{1987PASP...99.1214V} at Lick Observatory in California, USA. We utilize the velocities and errors reported in \citet{2014ApJS..210....5F}. Velocity errors from the Lick planet search include instrument stability, but do not account for stellar jitter. These are relative velocities (the star compared to itself) and hence cannot be directly compared to other measurements without modeling an offset \citep{2016A&A...591A.146D}.

\section{Host Star Analysis}\label{sec:star}
We summarize constraints on the host star in Table~\ref{tab:prop}, the details of which we provide in in this section. 

\begin{deluxetable}{lccc}
\centering
\tabletypesize{\scriptsize}
\tablewidth{0pt}
\tablecaption{Properties of the host star \name. \label{tab:prop}}
\tablehead{\colhead{Parameter} & \colhead{Value} & \colhead{Source} }
\startdata
\multicolumn{3}{c}{Astrometry}\\
\hline
$\alpha$.  & 07:49:55.06 & \emph{Gaia} DR2\\
$\delta$. & +27:21:47.5 & \emph{Gaia} DR2 \\
$\mu_\alpha$ (mas\,yr$^{-1}$)& -10.027$\pm$0.085 & \emph{Gaia} DR2\\
$\mu_\delta$  (mas\,yr$^{-1}$) & -11.314$\pm$0.049 & \emph{Gaia} DR2\\
$\pi$ (mas) & 44.607$\pm$0.044 & \emph{Gaia} DR2\\
\hline
\multicolumn{3}{c}{Photometry}\\
\hline
G$_{Gaia}$ (mag) & 6.7183$\pm$0.0005 & \emph{Gaia} DR2\\
BP$_{Gaia}$ (mag) & 7.0919$\pm$0.0021 & \emph{Gaia} DR2\\
RP$_{Gaia}$ (mag) & 6.2322$\pm$0.0022 & \emph{Gaia} DR2\\
B$_T$ (mag) & $7.749\pm0.016$ & Tycho-2 \\
V$_T$ (mag) & $6.987\pm0.010$ & Tycho-2\\
J (mag) & $5.624\pm0.043$ &  2MASS\\
H (mag) & $5.359 \pm0.026$  & 2MASS\\	
Ks (mag) & $5.258\pm0.016$ & 2MASS\\
W1 (mag) & $5.246\pm0.178$ & ALLWISE\\
W2 (mag)& $5.129\pm0.087 $ & ALLWISE\\
W3 (mag)& $5.297\pm0.016$ & ALLWISE\\ 
W4 (mag)& $5.163 \pm0.031$ & ALLWISE\\ 
\hline
\multicolumn{3}{c}{Kinematics \& Position}\\
\hline
RV$_{\rm{Bary}}$ (km\, s$^{-1}$) & -15.81$\pm$0.10 & This paper\\
U (km\, s$^{-1}$) & 13.66$\pm$0.09 & This paper\\
V (km\, s$^{-1}$) & 2.42$\pm$0.02 & This paper\\
W (km\, s$^{-1}$) & -7.75$\pm$0.04 & This paper\\
X (pc) & -19.89$\pm$0.02 & This paper\\
Y (pc) & -4.697$\pm$0.005 & This paper\\
Z (pc) & 9.164$\pm$0.091 & This paper\\
\hline
\multicolumn{3}{c}{Physical Properties}\\
\hline
$P_{\rm{rot}}$ (days) &  $6.45\pm0.05$ & This paper\\
$L_X/L_{\rm{bol}}$ & $(9.1\pm2.7)\times10^{-5}$& This paper\\
$\log R'_{\rm{HK}}$ & $-4.39\pm0.05$ & This paper\\
\vsini (km\, s$^{-1}$) & $ 7.3\pm0.3 $ & This paper\\
$i_*$ ($^\circ$) & $ >71$ & This paper\\
\fbol\,(erg\,cm$^{-2}$\,s$^{-1}$)& ($4.823\pm0.12)\times10^{-8}$ & This paper\\ 
T$_{\mathrm{eff}}$ (K) & $5640 \pm 74$ & This paper\\
M$_\star$ (M$_\odot$) & $ 0.99\pm0.03 $ & This paper \\
R$_\star$ (R$_\odot$) &  $0.912 \pm 0.034$ & This paper \\
L$_\star$ (L$_\odot$) & $0.753 \pm 0.026$ & This paper \\
$\rho_\star$ ($\rho_\odot$) & $1.3\pm0.15$ & This paper \\
Age (Myr) & $414 \pm 23$ & \citet{UrsaMajorAge} \\
\enddata
\end{deluxetable}

\subsection{Membership to Ursa Major and Age}\label{sec:member}

The Ursa Major Group (UMaG) has long been proposed as a kinematically similar grouping of stars \citep[e.g.,][]{1869RSPS...18..169P, 1921MeLuS..26....3R, 1965Obs....85..104E} centered on several of the stars comprising the Ursa Major constellation. While UMaG has a clear core of members that are homogeneous in kinematics and color-magnitude-diagram position \citep{1993AJ....105..226S, King2003}, many associations with large spatial distributions have turned out to be larger star-formation events with multiple ages \citep[e.g., Sco-Cen and Taurus-Auriga,][]{rizzuto2011, Kraus2017a}. Further, the spatial spread of UMaG members outside the core leads to a large number of interloping stars with similar Galactic orbits, but different ages. \citet{2017A&A...597A..33T} found that $\simeq \sfrac{2}{3}$ of UMaG members have similar chemical compositions, suggesting either multiple stellar populations or a large fraction of contaminants in the membership list. Thus, to be useful for age-dating \name, we need to establish its association to the core members of UMaG using both kinematics and independent metrics (e.g., rotation and abundances). 

UMaG has recent age estimates ranging from 390-530\,Myr \citep[e.g.,][]{2015ApJ...804..146D, 2015ApJ...807...58B}. Direct radius measurements from long-baseline interferometry for the A stars in UMaG point toward a common age of $\tau = 414 \pm 23$ Myr \citep{UrsaMajorAge}. We adopt this measurement as the cluster age for analyses in this paper.

\name\ was first identified as a possible member of UMaG by \citet{Gaidos1998} based on its kinematics and X-ray luminosity, and has since been included as a candidate or likely member in multiple analyses \citep[e.g.,][]{King2003, 2008MNRAS.384..173F, 2018ARep...62..502V}. The spatial and kinematic definition of UMaG was most recently updated by \citet{BanyanSigma}, who identified central values of $(X,Y,Z) = (-7.5,+9.9,+21.9)$ pc and $(U,V,W) = (+14.8,+1.8,-10.2)$~\kms, along with full covariance matrices for these parameters. These are marginally consistent with the central values from \citet{2010AJ....139..919M} of $(U, V, W) = (+15.0, 2.8, -8.1)$ $\pm$ (0.4, 0.7, 1.0)\,\kms. The \gaia\ DR2 proper motion, parallax, and our radial velocity for \name\ give $(U,V,W) = (+13.66,+2.42,-7.75)$ \kms. Following the method from \citet{BanyanSigma}\footnote{\url{https://github.com/jgagneastro/banyan_sigma_idl}}, we calculate a membership probability of $P_{mem} = 99.98\%$ using the \citet{BanyanSigma} space velocity for UMaG, and $P_{mem} = 95\%$ using the space velocity from \citet{2010AJ....139..919M}. 

\name's Galactic position of $(X, Y, Z) = (-19.89, -4.70, 9.16)$\,pc is 23\,pc from the core of UMaG at $(X, Y, Z)=(-7.5, 9.9, 21.9)$\,pc. While the core members of UMaG are within $\simeq$4\,pc of the core, more than half of known members $>$20\,pc away \citep{2002A&A...381..446M}. The large-scale velocity dispersion is $\simeq1$\,\kms\ ($\simeq$1\,pc/Myr) and the age is $\tau \sim 400$\,Myr; members could easily be spread over $>$100\,pc in $Y$ (where orbits can freely diverge) and a still substantial distance even in $X$ and $Z$ (where epicyclic motion prevent the spatial distribution from broadening to the same degree). \citet{2010AJ....139..919M} and \citet{2016ApJ...818....1S} argue that the measured dispersion of $\simeq1$\,\kms\ is mostly an artifact of including spectroscopic binaries in the sample, and that the true dispersion is smaller. However, given the age of the cluster, \name\ only needs to have a velocity difference of $\simeq$0.06\kms\ to explain a 23\,pc separation. It is more likely that the velocity difference is larger and \name\ was evaporated from the cluster core in the last 100\,Myr. 

The photospheric lithium abundance provides an age and membership diagnostic that is independent of the 6D position-velocity phase space of \name. Li is destroyed at temperatures common to the cores of stars ($\sim2.5\times10^6$\,K), which slowly depletes surface Li at a rate that depends on the core-surface transport efficiency (e.g., convection). While there is significant scatter within a single age group \citep[e.g.,][]{2015MNRAS.449.4131S}, there is still a shift in the average A(Li)-\teff\ sequence with age. 

\begin{figure}[tb]
    \centering
    \includegraphics[width=0.49\textwidth]{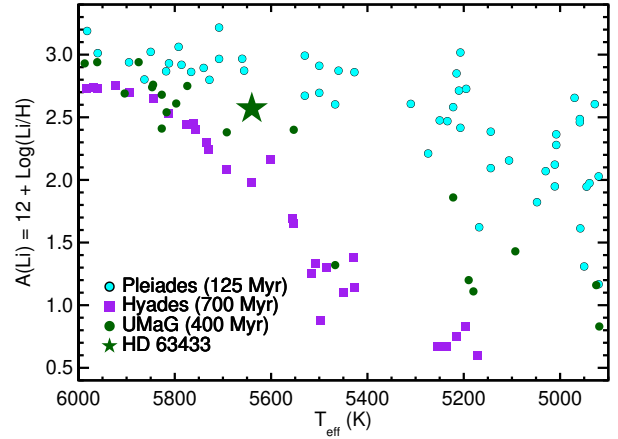}
    \caption{A(Li) sequence as a function of \teff\ for Hyades (purple), Pleiades (cyan) and UMaG (green). UMaG lands in between the two clusters, as expected for its intermediate age. \name\ is shown as a green star; its A(Li) abundance is within the expected sequence for UMaG between Hyades and Pleiades. }
    \label{fig:lithium}
\end{figure}

We compared the A(Li) abundance of \name\ from \citet{Ramirez2012} to A(Li) for members of Pleiades \citep[125\,Myr; ][]{Dahm2015} and Hyades \citep[700\,Myr; ][]{Martin2018}. A(Li) measurements for the Pleiades were taken from \citet{2018A&A...613A..63B}, and values for the Hyades from \citet{2016ApJ...830...49B}. We also considered UMaG members that were confirmed using kinematics and chromospheric activity by \citet{King2003}. We retrieved A(Li) from \citet{2005PASP..117..911K}, \citet{Ramirez2012}, \citet{2018A&A...614A..55A}, and the Hypatia catalog \citep{2014AJ....148...54H}. \name\ has A(Li) between that of Hyades and Pleiades stars of similar \teff, and is consistent with the core members of UMaG, furthering the case for membership (Figure~\ref{fig:lithium}).

Stellar rotation provides an additional check on the age and membership of \name. Once on the main sequence, young stars lose angular momentum with time, decreasing their rotation periods. After 100-600\,Myr, Sun-like stars eventually converge to a sequence \citep{vanSaders2016, Douglas2016}. \citet{Angus2015} used this information to provide a calibration between $B-V$, $P_{\rm{rot}}$, and age, which predicts a rotation period of $6.9\pm0.4$\,d for \name\ if it is a member of UMaG.

\begin{figure*}[t]
    \centering
    \includegraphics[width=0.96\textwidth]{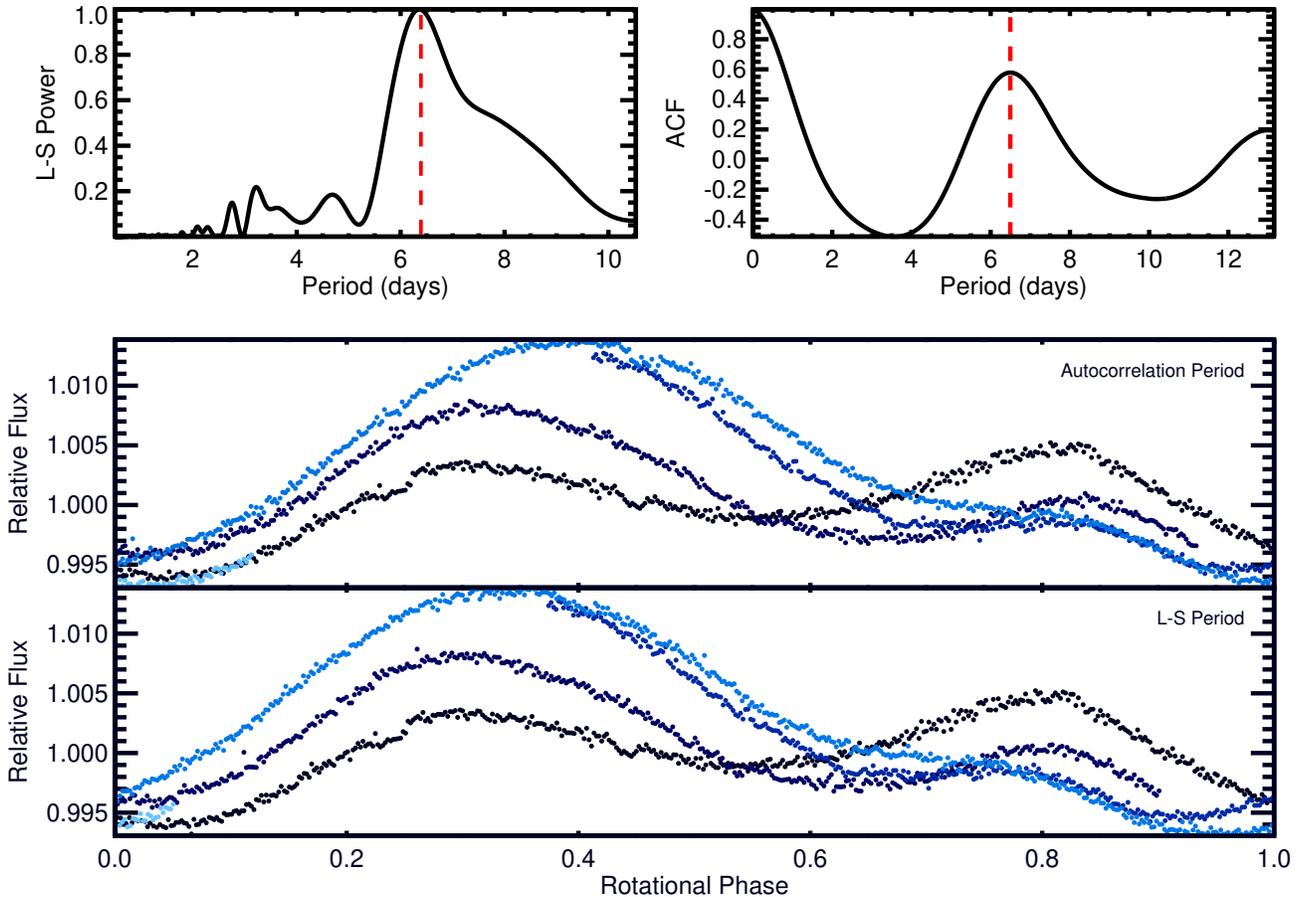}
    \caption{Diagnostic plot of our rotation period estimate for \name. The top two panels show the Lomb-Scargle (left) and autocorrelation (right) power, with dashed lines indicating the assigned period. The bottom two panels show the \tess\ light curve phase-folded to the two periods from Lomb-Scargle (bottom) and the autocorrelation function (middle).}
    \label{fig:prot}
\end{figure*}

We estimated the rotation period of \name\ and other likely UMaG members from their {\it TESS} or \ktwo\ light curves using a combination of the Lomb-Scargle periodogram following \citet{LombScargle} and the autocorrelation function as described in \citet{McQuillan2013}. For this, we used the simple-aperture-photometry (SAP) lightcurves, as PDCSAP curves tend to have long-term signals removed or suppressed \citep[e.g.,][]{Smith2012, VanCleve2016}. For \name, this yielded a period estimate of 6.45\,d (Figure~\ref{fig:prot}). The Lomb-scargle power is relatively broad, although the power is high and bootstrap resampling of the light curve suggest on this period of $\lesssim$3\%. Our derived period is also consistent with the literature estimate of 6.46$\pm$0.01\,d from \citet{2000AJ....120.1006G}. Rotation periods for other likely members are listed in Table~\ref{tab:UMaG_rot}. We note that because of the narrow window provided by \tess\ photometry for many UMaG members ($<$30\,d), our estimates are subject to aliasing (i.e., true periods may be double or half the assigned value) separate from the smaller formal errors ($\simeq$3\%). 

\begin{deluxetable}{lrrrrr}
\centering
\tabletypesize{\scriptsize}
\tablewidth{0pt}
\tablecaption{Rotation Periods for likely Members of UMaG \label{tab:UMaG_rot}}
\tablehead{\colhead{Object} & \colhead{TIC} & \colhead{RA} & \colhead{Dec} & \colhead{Bp-Rp} & \colhead{Prot} \\
\colhead{} & \colhead{} & \colhead{(deg)} & \colhead{(deg)} & \colhead{(mag)} & \colhead{(days)} }
\startdata
HD109011  & 316331312 & 187.82883 & 55.11897  & 1.193 & 8.40  \\
HD109647  & 224305606 & 188.96370 & 51.22148  & 1.180 & 4.57  \\
HD109799  & 60709182  & 189.42616 & -27.13888 & 0.468 & 0.79  \\
HD110463  & 99381773  & 190.43551 & 55.72467  & 1.165 & 12.07 \\
HD11131   & 24910401  & 27.34729  & -10.70362 & 0.804 & 9.16  \\
HD111456  & 142277151 & 192.16436 & 60.31973  & 0.665 & 1.47  \\
HD11171   & 24910738  & 27.39626  & -10.68641 & 0.456 & 0.76  \\
HD113139A & 229534764 & 195.18163 & 56.36633  & 0.515 & 0.73  \\
HD115043  & 157272202 & 198.40420 & 56.70827  & 0.789 & 5.53  \\
HD147584  & 362747897 & 247.11725 & -70.08440 & 0.725 & 8.22  \\
HD165185  & 329574145 & 271.59883 & -36.01979 & 0.780 & 5.90  \\
HD180777  & 235682463 & 287.29116 & 76.56050  & 0.433 & 0.77  \\
HD238224  & 159189482 & 200.84697 & 57.90606  & 1.657 & 12.13 \\
HD26923   & 283792884 & 63.87000  & 6.18686   & 0.747 & 5.78  \\
HD38393   & 93280676  & 86.11580  & -22.44838 & 0.720 & 12.89 \\
HD59747   & 16045498  & 113.25242 & 37.02985  & 1.059 & 7.94  \\
HD63433   & 130181866 & 117.47942 & 27.36318  & 0.860 & 6.39  \\
HD72905   & 417762326 & 129.79877 & 65.02091  & 0.800 & 4.95  \\
HD95650   & 97488127  & 165.65976 & 21.96714  & 2.015 & 13.75 \\
\enddata
\end{deluxetable}

\begin{figure}[tb]
    \centering
    \includegraphics[width=0.47\textwidth]{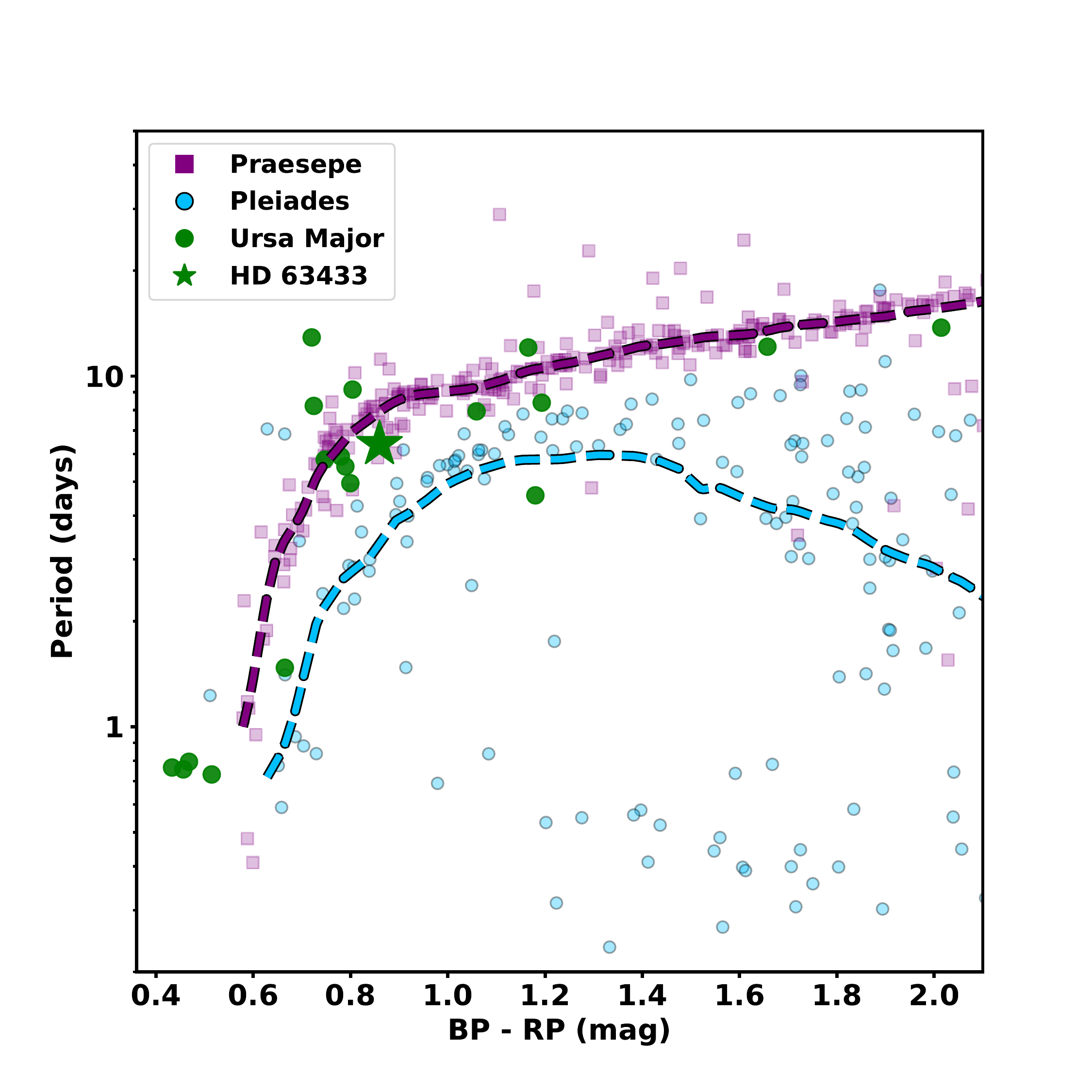}
    \caption{Rotation period versus \gaia\ $BP-RP$ color for members of the $\simeq$700\,Myr Praesepe clusters (purple), $\simeq125$\,Myr Pleiades cluster (cyan), and $\simeq$400\,Myr UMaG (green). \name\ is shown as a green star. For clarity, we also show a running median excluding stars with $<$0.2\,day rotation periods for Praesepe and Pleiades (dashed lines). While there is significant scatter and overlap between all three distributions, \name\ matches the sequence expected for UMaG's assigned age.} 
    \label{fig:rots}
\end{figure}

As with A(Li), we compare our UMaG rotation periods to those from the older Praesepe cluster \citep[from ][]{Douglas:2019} and younger Pleiades cluster \citep[from ][]{Rebull2016}. We show the results in Figure~\ref{fig:rots}. There is significant scatter in each sequence (e.g., from binarity and non-member interlopers). The  UMaG sequence lands just below the Praesepe sequence in period space, and most members are above (longer periods) than the Pleiades sequence. As expected, \name\ follows the overall trend for UMaG. The rotation sample considered here is small, and a more complete accounting of UMaG membership is needed to explore the overlap between Praesepe and UMaG rotation sequences.

Altogether, the available evidence confirms the age and membership of \name. For all analyses in the rest of the paper, we adopt the cluster age (414$\pm$23\,Myr) as the age of \name. 

\subsection{Spectral Energy Distribution Fit}\label{sec:SED}

We fit \name's spectral-energy-distribution using available photometry, our Goodman optical spectrum (Section~\ref{sec:goodman}), and spectral templates of nearby stars \citep[e.g.,][]{Rayner2009, Falcon2011}. To this end, we followed the basic methodology of \citet{Mann2015b}. The procedure gives precise (1-5\%) estimates of \fbol\ from the integral of the absolutely calibrated spectrum, $L_*$ from \fbol\ and the \gaia\ distance, and \teff\ from comparing the calibrated spectrum to atmospheric models. This method reproduces angular diameter measurements from long-baseline optical interferometry \citep[e.g.,][]{von-Braun:2012lq}. As a check, the same procedure also provides an estimate of $R_*$ from the infrared-flux method \citep{Blackwell1977}, i.e., the ratio of the absolutely calibrated spectrum to the model spectrum is $R_*^2/D^2$ \citep[also see Equation 1 of][]{Cushing2008}.

We first combined our template and observed spectra with Phoenix BT-SETTL models \citep{Allard2011} to cover wavelength gaps and regions of high telluric contamination and assumed a Rayleigh-Jeans law redward of where the models end (25-30\um). To absolutely calibrate the combined spectra, we used literature optical and NIR photometry from the Two-Micron All-Sky Survey \citep[2MASS; ][]{Skrutskie2006}, the Wide-field Infrared Survey Explorer \citep[WISE; ][]{allwise}, \gaia\ data release 2 \citep[DR2; ][]{Evans2018, GaiaDr2}, AAVSO All-Sky Photometric Survey \citep[APASS; ][]{apass}, Tycho-2 \citep{Hog2000}, Hipparcos \citep{vanLeeuwen1997}, and the General Catalogue of Photometric Data \citep{Mermilliod1997}. To account for variability of the source, we assumed all photometry had an addition error of 0.02\,mag (for optical) or 0.01\,mag (for near-infrared), and filter zero-points are assumed to have errors of 0.015\,mag unless a value is given in the source. We then compared the literature photometry to synthetic magnitudes derived from combined spectrum using the relevant filter profiles and zero points \citep[e.g.,][]{Mann2015a, dr2_filter}. We assumed no reddening, as \name\ lands within the local bubble \citep{LocalBubble}. In addition to the overall scale of the spectrum, there are four free parameters of the fit that account for imperfect (relative) flux calibration of the spectra and both the model and template spectra used. 

\begin{figure}[t]
    \centering
    \includegraphics[width=0.49\textwidth]{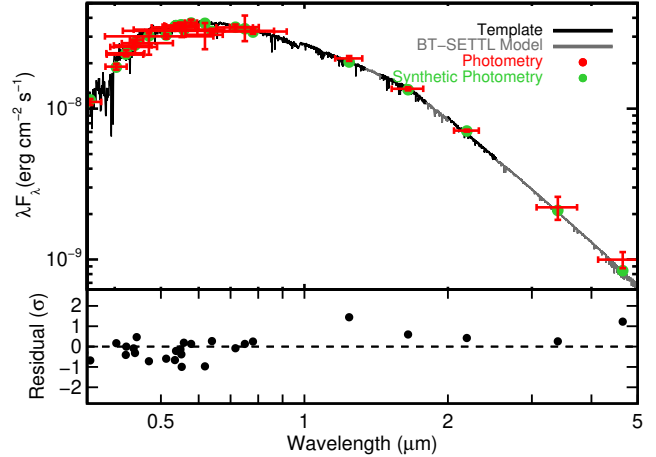}
    \caption{Best-fit spectral template and Goodman spectrum (black) compared to the photometry of \name. Grey regions are BT-SETTL models, used to fill in gaps or regions of high telluric contamination. Literature photometry is shown in red, with horizontal errors corresponding to the filter width and vertical errors to the measurement errors. Corresponding synthetic photometry is shown as green points. The bottom panel shows the residuals in terms of standard deviations from the fit.}
    \label{fig:sed}
\end{figure}

We show the best-fit spectrum and photometry in Figure~\ref{fig:sed}. There is no significant NIR excess seen out to $W4$, consistent with most stars at this age \citep{2008ApJ...679..720C}. Our joint fitting procedure yielded \teff=$5640\pm71$\,K, \fbol=$4.823\pm0.12 \times 10^{-8}$\,erg\,cm$^{-2}$\,s$^{-1}$, $L_*=0.753\pm0.026\,L_\odot$, and $R_*=0.912\pm0.034 R_\odot$. Our derived \teff\ is $<1\sigma$ consistent with literature determinations using high-resolution spectra \citep[5600--5700\,K; ][]{Baumann2010, Ramirez2012, Luck2017} and all parameters match our model interpolation below. 

Version 8 of the \tess\ Input Catalog \citep[TICv8; ][]{TIC2018, TIC2019} lists stellar parameters of \teff$=5693\pm153$\,K, [Fe/H]$=0.017\pm0.017$, $R_*=0.903\pm0.055R_\odot$, and $L_*=0.772\pm0.020L_\odot$. These are all in $\simeq1\sigma$ agreement with our SED-based parameters. 

\subsection{Spectroscopic Classification}

We derived spectral parameters from the TRES spectra of \name\ using the Spectral Parameter Classification (SPC) tool \citep{2012Natur.486..375B}. SPC cross correlates the observed spectrum against a grid of synthetic spectra based on Kurucz atmospheric models \citep{1993KurCD..13.....K}. \teff, \logg, bulk metallicity ([M/H]), and \vsini\ are allowed to vary as free parameters. This yielded \teff = $5705\pm 50$~K, \logg$ = 4.59 \pm 0.10$, [M/H]$= -0.09\pm 0.08$. 

We ran a similar analysis using the HARPS-N stacked spectrum with ARES/MOOG following \citet{Sousa2015}. Including empirical corrections from \citet{Sousa2011} and \citet{Mortier2014} yielded \teff=$5764\pm73$\,K, \logg$=4.65\pm0.12$ and [Fe/H]$=0.04\pm0.05$.

Both methods were consistent with our \teff\ derived using the SED, and the metallicity esitmates agrees with the established value for UMaG \citep[$-0.03\pm0.05$; ][]{2009A&A...508..677A}.

\subsection{Evolutionary-model parameters}

To determine the mass of \name, we used Mesa Isochrones and Stellar Tracks \citep[MIST; ][]{MIST1}. We compared all available photometry to the model-predicted values, accounting for errors in both the photometric zero-points and stellar variability as in Section~\ref{sec:SED}. We restrict the comparison to 300--600\,Myr and solar metallicity based on the properties of the cluster. We assumed Gaussian errors on the magnitudes but included a free parameter to describe underestimated uncertainties in the models or data. The best-fit parameters from the MIST models were $M_*=0.991\pm0.027M_\odot$, $R_*=0.895\pm0.021R_\odot$, \teff=5690$\pm$61\,K, and   $L_*=0.784\pm0.031L_\odot$. These are consistent with our other determinations, but we adopt our empirical $L_*$, \teff\ and $R_*$ estimates from the SED and only utilize the $M_*$ value from the evolutionary models in our analysis.

\subsection{Stellar Inclination}\label{sec:inc}

Using the combination of projected rotation velocity (\vsini), $P_{\rm{rot}}$, and $R_*$, we can estimate the stellar inclination ($i_*$), and hence test wheather the stellar spin and planetary orbit are consistent with alignment. In principle, this is done by estimating the $V$ term in \vsini\ using $V=2\pi R_*/P_{\rm{rot}}$, although in practice it requires additional statistical corrections, including the fact that we can only measure alignment projected onto the sky. To this end, we follow the formalism from \citet{2020AJ....159...81M}, which handles the hard barrier at $i_*>90^\circ$ by rewriting the relation in terms of $\cos(i)$.

We used our $P_{\rm{rot}}$ from Section~\ref{sec:member} estimated from the \tess\ light curve, and our $R_*$ derived in Section~\ref{sec:SED}. \name\ has \vsini\ measurements from a range of literature sources, with estimates from 7.0\,\kms\ \citep{2014MNRAS.444.3517M} to 7.7\,\kms\ \citep{Luck2017}. Our fit of the TRES spectra yielded a consistent estimate of \vsini\ $ = 7.3 \pm 0.3$\,\kms\ with a macroturbulent velocity of 4.2$\pm$1.2\,\kms\ and the SpecMatch run on the NRES spectra yielded 7.1$\pm$0.3\,\kms. We adopted 7.3$\pm$0.3\,\kms, which encompassed all estimates. 

The combined parameters yielded a equatorial velocity ($V$) of 7.16$\pm$0.29\,\kms\ and a lower limit for the inclination of $i_*>71^\circ$ at 68\% confidence and $i>68^\circ$ at 95\% confidence. This is consistent with the stellar rotation being aligned with the planetary orbits ($i\simeq90^\circ$).

\subsection{Limits on Bound, Spatially Resolved Companions}\label{sec:AO}

\name\ has adaptive optics or interferometric data spanning almost a decade, from 1999 \citep{WDS2001} to 2008 \citep{CIA2012b}. The deepest extant high-resolution imaging reported in the literature for \name\ was obtained with the NaCo instrument at the VLT on 2004 Jan 16 UT (Program 072.C-0485(A); PI Ammler). The observation consisted of a series of individual 35 sec exposures, totaling 980 sec in all, taken with the $K_s$ filter and with the central star behind a 0.7\arcsec\, opaque Lyot coronagraph. The results of these observations were reported by \citet{2016A&A...591A..84A}, who found no candidate companions within $\rho < 9\arcsec$. The detection limits were reported in a figure in that work, and achieved contrasts of $\Delta K_s \sim 7$ mag at 0.5\arcsec, $\Delta K_s \sim 9$ mag at 1\arcsec, $\Delta K_s \sim 11$ mag at 2\arcsec, and $\Delta K_s \sim 13$ mag at $\ge$3\arcsec. Given an age of $\tau \sim 400$ Myr, the evolutionary models of \citet{BHAC15} would imply corresponding physical limits of $M < 65 M_{Jup}$ at $\rho \sim 11$ AU, and $M < 50 M_{Jup}$ at $\rho > 22$ AU.

\cite{WDS2001} and \cite{CIA2012b} reported null detections at higher spatial resolution using speckle and long-baseline interferometry. These observations are consistent with the limits set by the lack of \gaia\ excess noise as indicated by the Renormalized Unit Weight Error \citep{GaiaDr2}. \name\ has $RUWE = 0.98$, consistent with the distribution of values seen for single stars. Based on a calibration of the companion parameter space that does induce excess noise, this corresponds to contrast limits of $\Delta G \sim 0$ mag at $\rho = 30$ mas, $\Delta G \sim 4$ mag at $\rho = 80$ mas, and $\Delta G \sim 5$ mag at $\rho \ge 200$ mas. The same evolutionary models would imply corresponding physical limits for equal-mass companions at $\rho \sim 0.7$ AU, $M < 0.4 M_{\odot}$ at $\rho \sim 1.8$ AU, and $M < 0.3 M_{\odot}$ at $\rho > 4.4$ AU.

Finally, the \gaia\ DR2 catalog \citep{GaiaDr2} did not report any comoving, codistant companions within $<1\degree$ of \name. \citet{2017AJ....153..257O} reported a comoving companion based on \gaia\ DR1 astrometry, but the DR2 parameters for the two stars are inconsistent with each other and the claimed companion is $>3^\circ$ away from \name. Given the \gaia\ catalog's completeness limit of $G \sim 20.5$ mag at moderately high galactic latitudes and sensitivity at $\rho > 3\arcsec$ \citep{Ziegler2018,2019A&A...621A..86B}, the absence of wide companions corresponds to a physical limit of $M < 0.05 M_{\odot}$ at $\rho > 66$ AU to $\rho \sim$80,000 AU.

\section{Light Curve Analysis}\label{sec:transit}

\subsection{Identification of the Transit Signals}

We identified a 7-day period planet candidate during a visual survey of the 2-minute cadence \tess\ Candidate Target List data \citep{TIC2018} via the light curve examining software LcTools\footnote{\url{https://sites.google.com/a/lctools.net/lctools/}} \citep{Schmitt:2019}, which initially was introduced by \citet{2015ApJ...813...14K}. Further inspection revealed two additional transits of similar depth and duration located at 1844.057760 and 1864.606371 TBJD (BJD-2457000), indicating the presence of a second planet candidate with a period of approximately 21-days. The first candidate was released as a \tess\ object of interest (TOI) from analysis of the SPOC light curve, the first on Feb 19, 2020 and the second on Feb 20, 2020, nearly simultaneous with our visual search. The SPOC data validation reports \citep{DVreports1, DVreports2} note a significant centroid offset for the outer planet, but the offset is not consistent with any nearby source and such offsets are common for young variable stars \citep{Newton2019}. This offset was likely owing to saturation (expected at $T=6.27$) which degrades the centroids in the row direction. 

\begin{figure}[tb]
    \centering
    \includegraphics[width=0.49\textwidth]{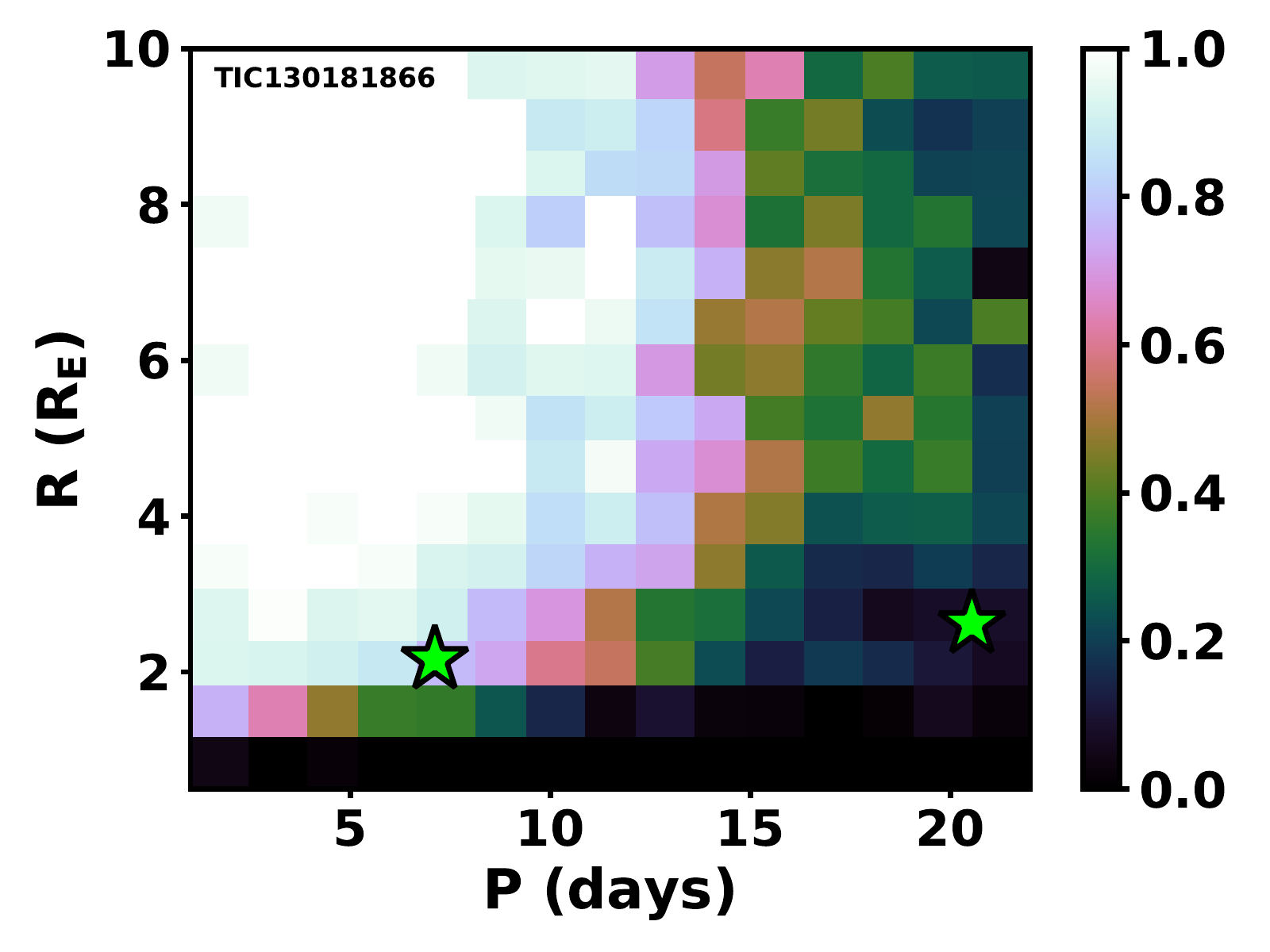}
    \caption{Completeness plot for the \tess\ light curve of \name\ based on the injection/recovery formalism described by \citet{Rizzuto2017}. Regions are color-coded by the fraction of systems recovered. The two real planets are shown as green stars. 
    }
    \label{fig:inject}
\end{figure}

We searched for additional planets using the notch filter, as described in \citet{Rizzuto2017}. We recovered the two planets identified above, but found no additional planet signals that passed all checks. Instead, we set limits on the existence of additional planets using an injection/recovery test, again following \citet{Rizzuto2017}. To briefly summarize, we generated planets using the \texttt{BATMAN} package following a uniform distribution in period, $b$, and orbital phase. Half of the planet radii are drawn from a uniform distribution and the other half from a $\beta$ distribution with coefficients $\alpha=2$ and $\beta=6$. We used this mixed distribution to ensure higher sampling around smaller and more common planets. We then detrended the light curve using the notch filter and searched for planets in the detrended curve using a Box-Least Squares (BLS) algorithm \citep{Kovacs2002}. The results are summarized in Figure~\ref{fig:inject}. We found that our search would be sensitive to $R_P\simeq 1 R_\oplus$ planets at periods of $<$5\,d and $R_P\simeq 2R_\oplus$ out to 15\,d. We required at least two transits to consider a signal recovered, but the light curve covers $<$30\,d, so most injected planets with $>15$\,d periods were not recovered. 

\subsection{MCMC Fit of the Transit and HARPS-N Velocities}

We fit the \tess\ photometry simultaneously with the HARPS-N velocities during transit (of the RM effect) using the \texttt{misttborn} (MCMC Interface for Synthesis of Transits, Tomography, Binaries, and Others of a Relevant Nature) fitting code\footnote{\url{https://github.com/captain-exoplanet/misttborn}} first described in \citet{Mann2016a} and expanded upon in \citet{MISTTBORN}. \texttt{misttborn} uses \texttt{BATMAN} \citep{Kreidberg2015} to generate model light curves and \texttt{emcee} \citep{Foreman-Mackey2013} to explore the transit parameter space using an affine-invariant Markov chain Monte Carlo (MCMC) algorithm. We did not include any of the other radial velocities in this analysis, especially given the complication of stellar activity, as they are not precise enough to detect the reflex motion owing to these small planets.

The standard implementation of \texttt{misttborn} fits for six parameters for each transiting planet: time of periastron ($T_0$), orbital period of the planet ($P$), planet-to-star radius ratio ($R_p/R_\star$), impact parameter ($b$), two parameters ($\sqrt{e}\sin{\omega}$ and $\sqrt{e}\cos{\omega}$) to characterize the orbital eccentricity ($e$) and argument of periastron ($\omega$), as well as three parameters related to the star: the stellar density ($\rho_\star$) and the linear and quadratic limb-darkening coefficients ($q_1$, $q_2$) following the triangular sampling prescription of \citet{Kipping2013}. 

To model stellar variations, \texttt{misttborn} includes a Gaussian Process (GP) regression module, utilizing the \texttt{celerite} code \citep{celerite}. For the GP kernel, we mostly followed \citet{celerite} and used a mixture of two stochastically driven damped simple harmonic oscillators (SHOs), at periods $P_{GP}$ (primary) and $0.5P_{GP}$ (secondary). In addition to the stellar rotation period ($\ln(P_{GP})$), the light curve kernel is characterized by a variability amplitude of the fundamental signal ($\ln{A_1}$), the decay timescale for the secondary signal ($\ln{Q_{2}}$, the quality factor), the difference between the quality factor the first and second signal ($\ln{\Delta Q} = Q_{1}-Q_{2}$), and a mix parameter (Mix) that describes the relative contribution of the two SHOs (where $A_1/A_2 = 1+e^{-\rm{Mix}}$). 

For the RM data, we used \texttt{misttborn} to fit for eight additional parameters for planet b. The primary parameters were the sky-projected spin-orbit misalignment ($\lambda$), the stellar rotation broadening (\vsini), and the intrinsic width of the Gaussian line profile of individual surface elements ($v_{\mathrm{int}}$), which approximates the combined effects of thermal, microturbulent, and macroturbulent broadening. We also included a quadratic polynomial fit to the out-of-transit variations in the radial velocity data ($\gamma$, $\dot{\gamma}$, and $\ddot{\gamma}$). We used a generic polynomial because the overall slope in the radial velocity curve is likely dominated by stellar activity, rather than a predictable sinusoidal curve induced by the planets. The last two parameters are the two limb-darkening coefficients ($q_{1,\mathrm{RM}}$ and $q_{2,\mathrm{RM}}$). These RM limb-darkening parameters were fit separately from those for the photometry because of differences in the HARPS-N and \tess\ wavelength coverage. From these parameters, we produced an analytic RM model following the methodology of \citet{Hirano11} and \citet{Addison14}. This model consists of an analytic function of $v\sin i$ and $v_{\mathrm{int}}$, multiplied by the flux drop owing to the transiting planet; we calculated the flux decrement needed for this model with \texttt{BATMAN}, following the same methodology as for the photometric light curves. We note that the GP described above is not used for the RM data, only the photometric curve. '

\begin{deluxetable}{llc}
\tablecaption{Parameters and Priors \label{tab:priors}}
\tablehead{\colhead{Parameter} & \multicolumn{2}{c}{Prior}\\
\colhead{} & \colhead{planet b} & \colhead{planet c} 
}
\startdata
$T_0$ (TJD)\tablenotemark{a} & $\mathcal{U}[1916.4,1916.5]$ & $\mathcal{U}[1842,1860]$ \\
$P$ (days) & $\mathcal{U}[0,\, 15]$ &  $\mathcal{U}[15,\, 30]$ \\
$R_P/R_{\star}$ & $\mathcal{U}[0, 1]$ & $\mathcal{U}[0,1]$ \\
$b$ & $\mathcal{U}[|b|<1+R_P/R_*]$ & $\mathcal{U}[|b|<1+R_P/R_*]$ \\
$\rho_{\star}$ ($\rho_{\odot}$) & \multicolumn{2}{c}{$\mathcal{N}[1.30, 0.15]$} \\
$q_{1,1}$ & \multicolumn{2}{c}{$\mathcal{N}[0.30,\,0.06] $}\\
$q_{2,1}$ & \multicolumn{2}{c}{$\mathcal{N}[0.37,\,0.05] $}\\
$\sqrt{e}\sin\omega$ & $\mathcal{U}[-1,1]$ & $\mathcal{U}[-1,\,1]$ \\
$\sqrt{e}\cos\omega$ & $\mathcal{U}[-1,1]$ & $\mathcal{U}[-1,\,1]$ \\
\hline
\vsini (km s$^{-1}$) & $\mathcal{N}[7.3,\,0.3] $ & \ldots \\
$\lambda$ ($^{\circ}$) & $\mathcal{U}[-180,180]$ & \ldots \\
$q_{1,\mathrm{RM}}$ & $\mathcal{N}[0.53,\,0.08]$ & \ldots \\
$q_{2,\mathrm{RM}}$ & $\mathcal{N}[0.39,\,0.06]$ & \ldots \\
$v_{\mathrm{int}}$ (km s$^{-1}$) & $\mathcal{N}[4.2\,1.2] $  & \ldots  \\
$\gamma_{1}$ (km s$^{-1}$) & $\mathcal{U}[-17,\, -14] $ & \ldots \\
$\dot{\gamma}$ (km s$^{-1}$) & $\mathcal{U}[-1,\, 1] $ & \ldots \\
$\ddot{\gamma}$ (km s$^{-1}$) & $\mathcal{U}[-1,\, 1] $ & \ldots \\
\hline
$\ln{P_{GP}}$ & \multicolumn{2}{c}{$\mathcal{U}[1,2] $ }\\
$\ln{A_{1}}$ &  \multicolumn{2}{c}{$\mathcal{U}[-\infty,\, 1] $ }\\
$\ln{(Q_0)}$ &  \multicolumn{2}{c}{$\mathcal{U}[0.5,\,\infty] $ }\\
$\ln{\Delta Q}$ &  \multicolumn{2}{c}{$\mathcal{U}[0,\infty] $ }\\
Mix & \multicolumn{2}{c}{$\mathcal{U}[-10,\, 10] $ } \\
\hline
\enddata
\tablecomments{$\mathcal{U}[X,Y]$ denotes a uniform prior limited to between $X$ and $Y$ and $\mathcal{N}[X,Y]$ a Gaussian prior with mean $X$ and standard deviation $Y$. }
\tablenotetext{a}{It is standard to report $T_0$ as the midtransit point for the {\it first} transit. However, for computational reasons in the RM fit, we restrict $T_0$ around the RM observations for planet b.}
\end{deluxetable}

We ran two separate MCMC chains, the first as described above, and the second with $e$ and $\omega$ locked at 0. For both chains, we ran the MCMC using 100 walkers for 250,000 steps including a burn-in of 20,000 steps. The autocorrelation time indicated that this was sufficient for convergence. We also applied Gaussian priors on the limb-darkening coefficients (both for \tess\ and the HARPS-N data) based on the values in \citet{Claret2011} and \citet{2015MNRAS.453.3821P}, with errors accounting for the difference between these two estimates (which differ by 0.05-0.07). For \vsini\ and $v_{\mathrm{int}}$, we used Gaussian priors of 7.3$\pm$0.3\,\kms\ and 4.2$\pm$1.2\,\kms\ based on analysis from Section~\ref{sec:inc} and the investigation of \citet{2014MNRAS.444.3592D}. For the fit with $e=0$, we applied Gaussian priors for the stellar density taken from our derived stellar parameters derived in Section~\ref{sec:SED}. All other parameters were sampled uniformly with physically motivated boundaries: $\sqrt{e}\sin{\omega}$ and $\sqrt{e}\cos{\omega}$ were restricted to $(-1,1)$, $|b|<1+R_p/R_s$, and $T_0$ to the time period sampled by the data. The GP mix parameter was restricted to be between -10 and 10. We let the linear and quadratic terms of the radial velocity curve in the RM data float, as this is produced by some combination of actual reflex motion of the star owing to the two planets, and stellar activity of this young, relatively rapidly rotating star. The full list of fit parameters, priors, and imposed limits are given in Table~\ref{tab:priors}.

\begin{figure*}[htp]
    \centering
    \includegraphics[width=\textwidth, trim=0 0 0 0,clip]{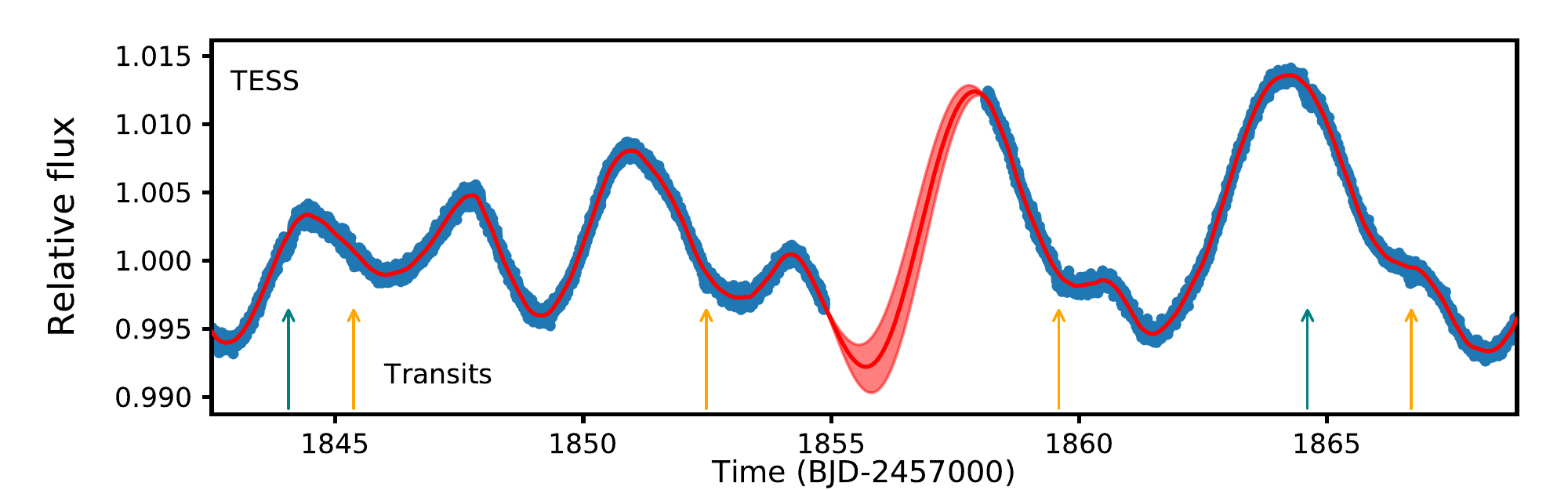}\\
    \includegraphics[width=\textwidth, trim=0 0.7cm 0 0,clip]{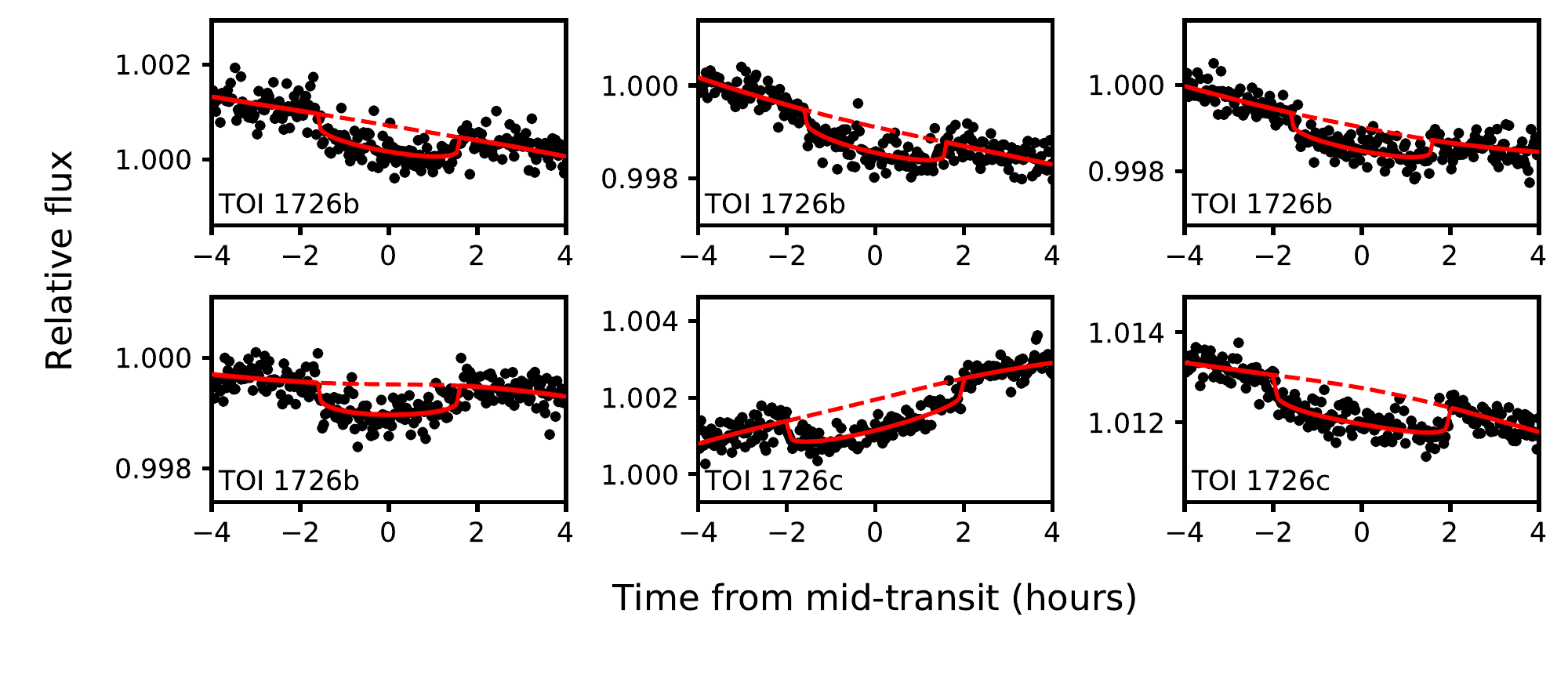}\\
    \caption{{\it TESS} light curve of \name. The top panel shows the PDCSAP curve (blue) after filtering out outliers, as well as our best-fit GP model (red). The locations of the transits are shown with arrows along the x-axis, red for planet b and teal for c. The bottom set of six panels shows the six individual transit events centered on the midtransit time with the best-fit model (red solid) and the best-fit model for the out-of-transit variability only (the GP; red dashed). Note that the x-axis scale and range of all individual transits are the same (hours from midtransit), but differ from the top panel (days), and y-axis ranges vary. }
    \label{fig:LCplot}
\end{figure*}

\begin{figure*}[htb]
    \centering
    \includegraphics[width=0.98\textwidth,trim=0 0 50 0, clip]{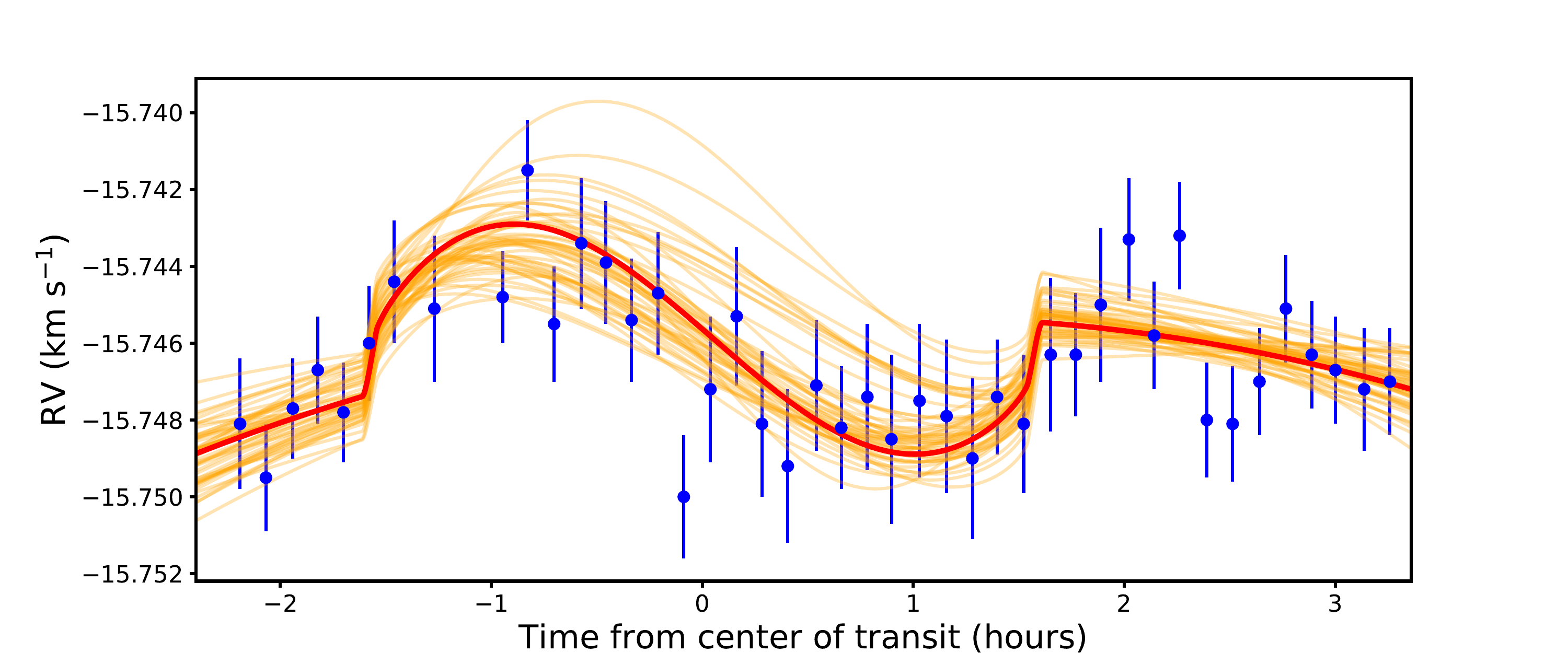}
    \caption{HARPS-N velocities and errors (blue points) compared to our best-fit RM model (red) and 50 random solutions drawn from the MCMC posteriors (orange). The orbit is clearly prograde, and the data favors a low value of $\lambda$ (larger amplitude), but a wide range of $\lambda$ values are allowed by the data.}
    \label{fig:rm}
\end{figure*}

\begin{figure*}[p]
    \centering
    \includegraphics[width=0.47\textwidth]{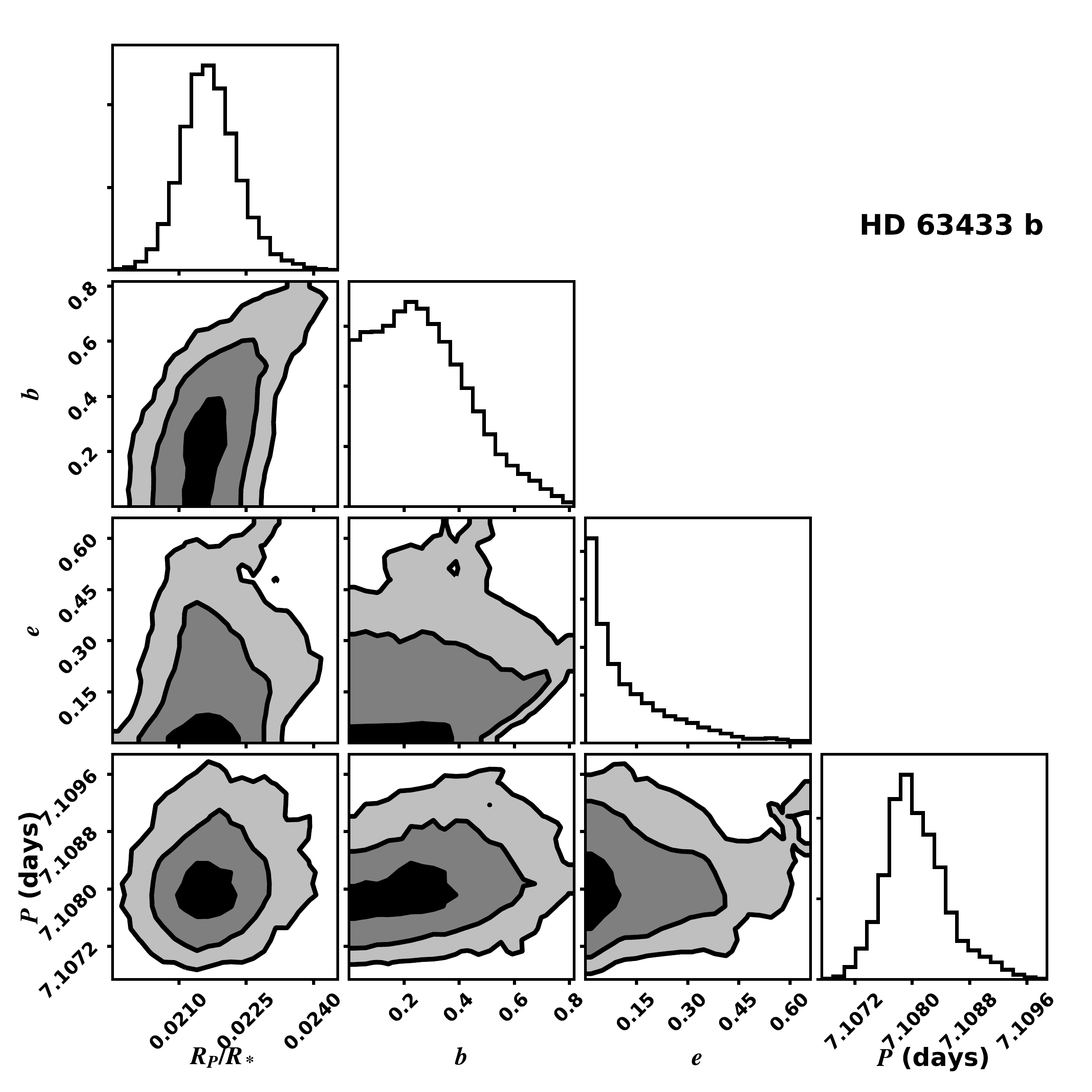}
    \includegraphics[width=0.47\textwidth]{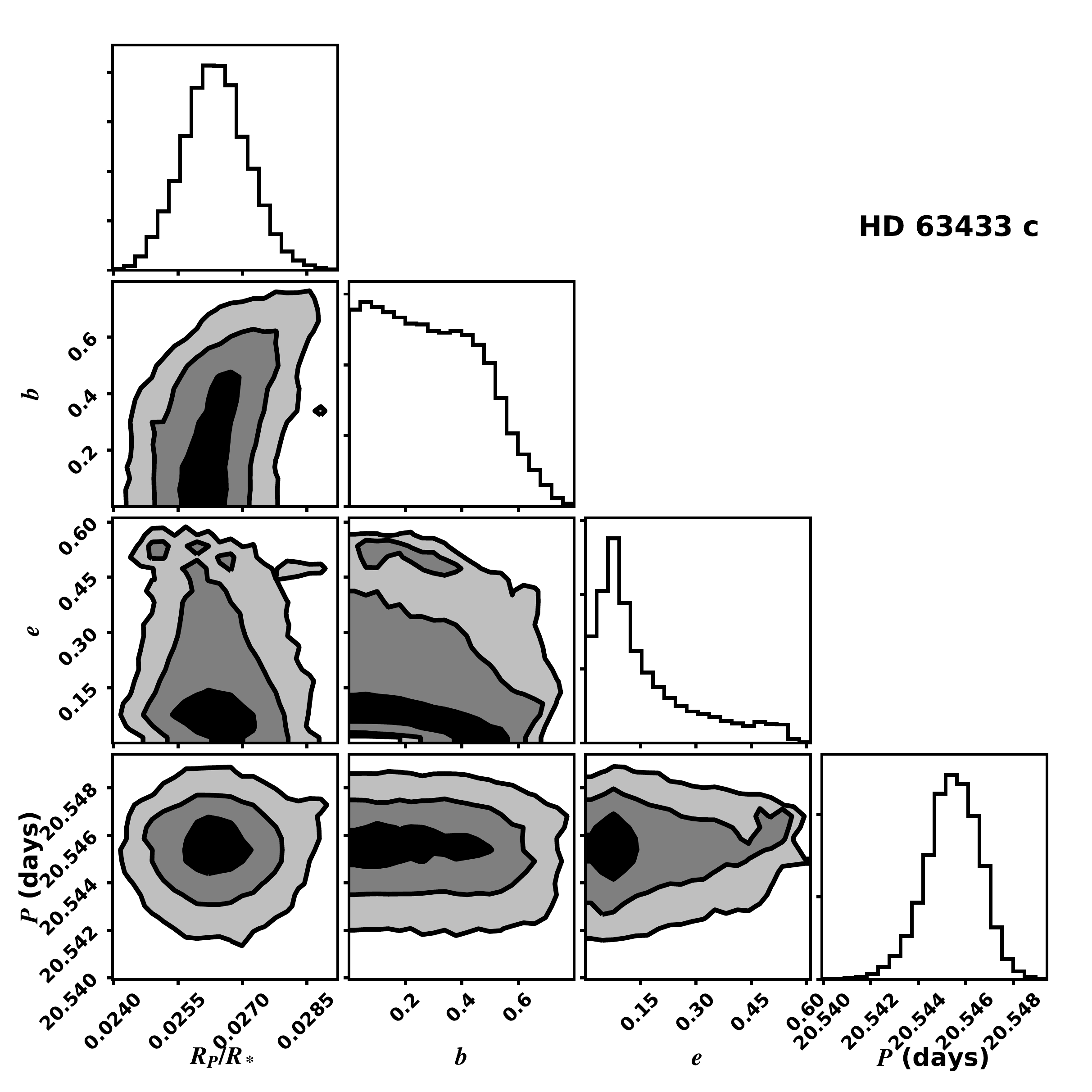}
    \includegraphics[width=0.47\textwidth]{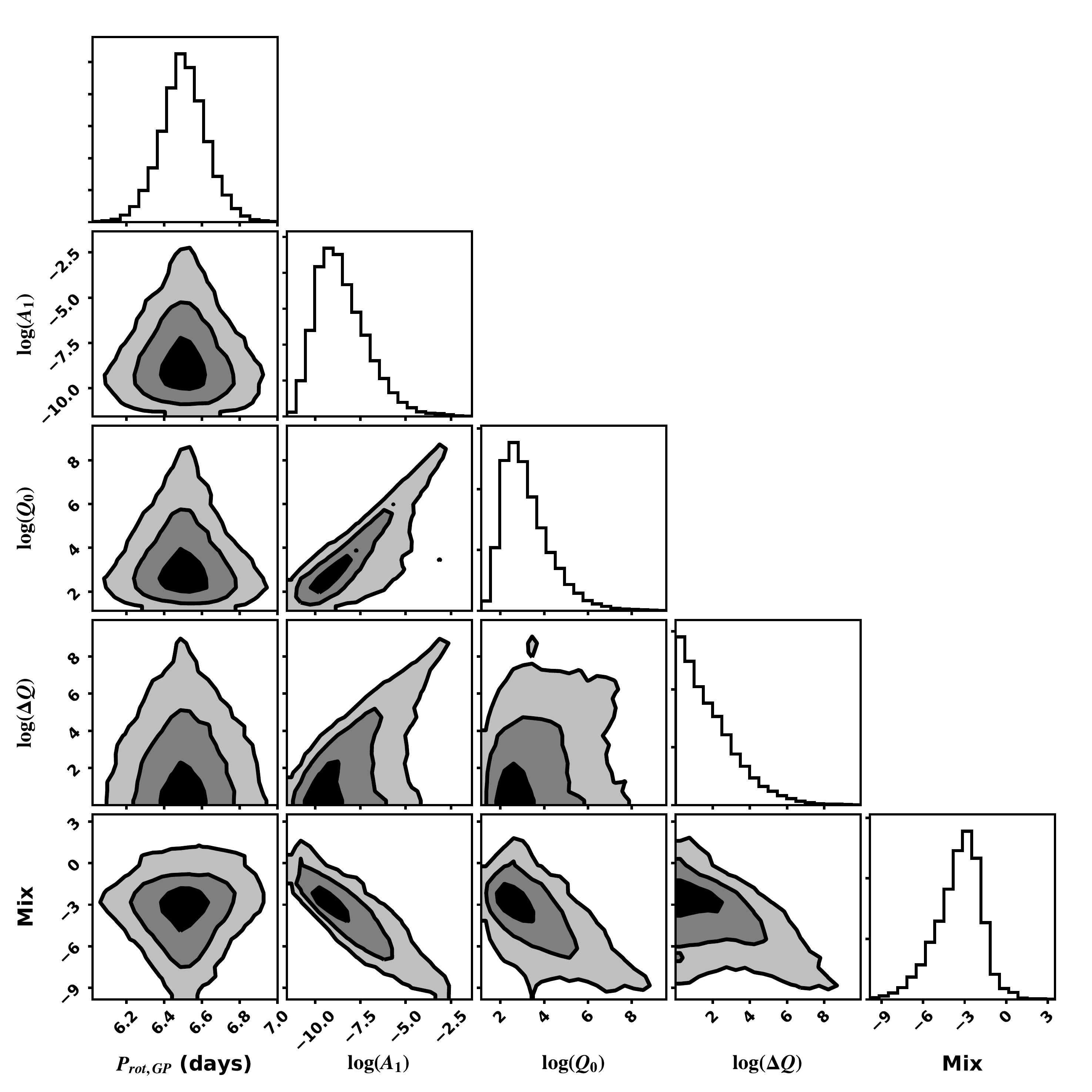}
    \includegraphics[width=0.47\textwidth]{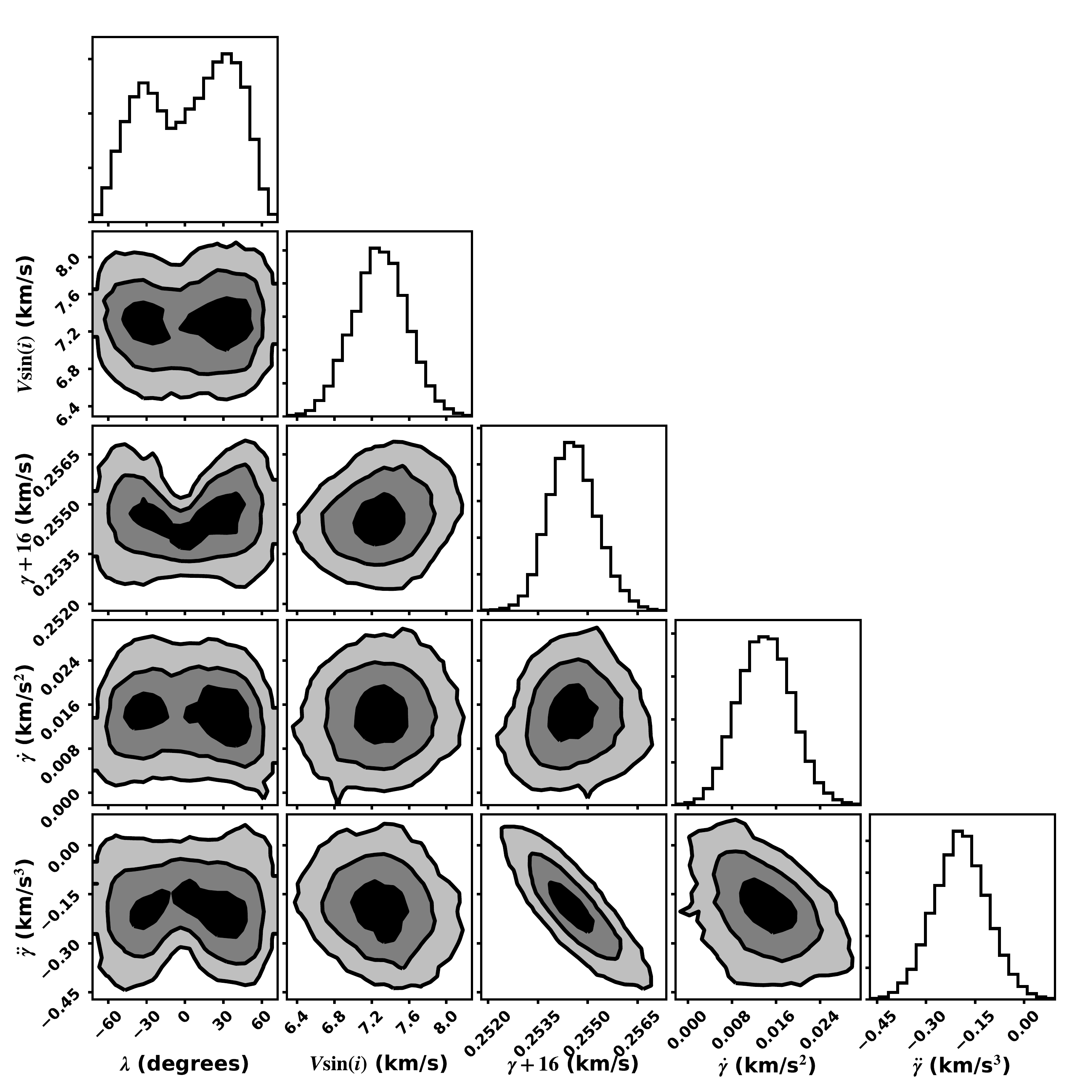}
    \caption{Posterior density and correlations for a subset of the parameters for planet b (top left), planet c (top right), the GP modelling (bottom left) and the RM fit (bottom right). A small percentage ($<1\%$) of points are cut off the plot edges for clarity. All parameters are fit simultaneously, but are shown as separate panels for clarity. We show physical parameters here rather than fitted ones, e.g., MCMC explores $\sqrt{e}\cos(\omega)$ and $\sqrt{e}\sin(\omega)$ while $e$ is shown here. See Section~\ref{sec:transit} for more details.}
    \label{fig:transcorner}
\end{figure*}

The resulting fit light curve is shown in Figure \ref{fig:LCplot} with the velocity curve in Figure~\ref{fig:rm}. The best-fitting model and derived parameters, along with 68\% credible intervals are listed in Table \ref{tab:transfit}. Figures \ref{fig:transcorner}a and b show partial posteriors for the MCMC-fit parameters\footnote{Trimmed posteriors for all parameters are available at \url{https://github.com/awmann/THYME3_HD63433}}. 

\begin{deluxetable*}{lccccccc}
\tabletypesize{\scriptsize}
\tablewidth{0pt}
\tablecaption{Transit-Fit Parameters. \label{tab:transfit} }
\tablehead{\colhead{Parameter} & \colhead{Planet b} &  \colhead{Planet c} & \colhead{Planet b} &   \colhead{Planet c} \\ 
\colhead{} &  \multicolumn{2}{c}{$e$,$\omega$ fixed} &\multicolumn{2}{c}{$e$,$\omega$ free}
 }
\startdata
\multicolumn{5}{c}{Transit Fit Parameters} \\
\hline
$T_0$ (TJD) &  $1916.4526^{+0.0032}_{-0.0027}$   &  $1844.05799^{+0.00084}_{-0.00087}$   &  $1916.4533^{+0.0037}_{-0.0027}$   &  $1844.05791^{+0.0008}_{-0.00076}$  \\
$P$ (days) &  $7.10793^{+0.0004}_{-0.00034}$   &  $20.5453^{+0.0012}_{-0.0013}$   &  $7.10801^{+0.00046}_{-0.00034}$   &  $20.5455 \pm 0.0011$  \\
$R_P/R_{\star}$ &  $0.02161 \pm 0.00055$   &  $0.02687 \pm 0.0007$   &  $0.02168^{+0.00065}_{-0.00058}$   &  $0.02637^{+0.00077}_{-0.00074}$  \\
$b$ &  $0.18^{+0.17}_{-0.13}$   &  $0.512^{+0.063}_{-0.033}$   &  $0.26^{+0.19}_{-0.17}$   &  $0.29^{+0.21}_{-0.2}$  \\
$\rho_{\star}$ ($\rho_{\odot}$) & \multicolumn{2}{c}{ $1.293^{+0.079}_{-0.2}$  } & \multicolumn{2}{c}{ $1.3 \pm 0.14$  } \\
$q_{1,1}$ & \multicolumn{2}{c}{ $0.302^{+0.058}_{-0.057}$  } & \multicolumn{2}{c}{ $0.299^{+0.06}_{-0.055}$  } \\
$q_{2,1}$ & \multicolumn{2}{c}{ $0.369 \pm 0.048$  } & \multicolumn{2}{c}{ $0.372 \pm 0.048$  } \\
$\sqrt{e}\sin\omega$ & 0 (fixed) & 0 (fixed) &  $-0.08^{+0.11}_{-0.13}$   &  $0.09^{+0.13}_{-0.18}$  \\
$\sqrt{e}\cos\omega$ & 0 (fixed) & 0 (fixed) &  $0.01^{+0.35}_{-0.37}$   &  $0.09^{+0.39}_{-0.43}$  \\
\hline
\multicolumn{5}{c}{Rossiter-McLaughlin Parameters}\\
\hline
$v\sin i_{\star}$ (km s$^{-1}$) &  $7.28 \pm 0.29$   & \nodata &  $7.3^{+0.29}_{-0.3}$   & \nodata\\
$\lambda$ ($^{\circ}$) &  $1.0^{+41.0}_{-43.0}$   & 0 (fixed) &  $8.0^{+33.0}_{-45.0}$   & 0 (fixed)\\
$q_{1,\mathrm{RM}}$ & \multicolumn{2}{c}{ $0.534^{+0.08}_{-0.081}$  } & \multicolumn{2}{c}{ $0.524^{+0.084}_{-0.078}$  } \\
$q_{2,\mathrm{RM}}$ & \multicolumn{2}{c}{ $0.388^{+0.056}_{-0.059}$  } & \multicolumn{2}{c}{ $0.388^{+0.057}_{-0.059}$  } \\
$v_{\mathrm{int}}$ (km s$^{-1}$) &  $4.3 \pm 1.1$   & \nodata &  $4.2 \pm 1.2$   & \nodata\\
$\gamma_{1}$ (km s$^{-1}$) &  $-15.74559^{+0.0007}_{-0.00064}$   & \nodata &  $-15.74542^{+0.00075}_{-0.0007}$   & \nodata\\
$\dot{\gamma}$ (km s$^{-1}$ day$^{-1}$) &  $0.0144^{+0.0049}_{-0.0048}$   & \nodata &  $0.014^{+0.0048}_{-0.0049}$   & \nodata\\
$\ddot{\gamma}$ (km s$^{-1}$ day$^{-2}$) &  $-0.184 \pm 0.078$   & \nodata &  $-0.198^{+0.082}_{-0.083}$   & \nodata\\
\hline
\multicolumn{5}{c}{Gaussian Process Parameters}\\
\hline
$\log(P_{GP})$ & \multicolumn{2}{c}{ $1.872^{+0.018}_{-0.017}$  } & \multicolumn{2}{c}{ $1.872 \pm 0.017$  } \\
$\log(A_{1})$ & \multicolumn{2}{c}{ $-8.5^{+1.7}_{-1.2}$  } & \multicolumn{2}{c}{ $-8.6^{+1.6}_{-1.1}$  } \\
$\log(\Delta Q)$ & \multicolumn{2}{c}{ $1.6^{+1.9}_{-1.1}$  } & \multicolumn{2}{c}{ $1.6^{+1.9}_{-1.2}$  } \\
$\log(Q_0)$ & \multicolumn{2}{c}{ $3.07^{+1.42}_{-0.85}$  } & \multicolumn{2}{c}{ $3.03^{+1.3}_{-0.82}$  } \\
Mix & \multicolumn{2}{c}{ $-3.5^{+1.3}_{-1.9}$  } & \multicolumn{2}{c}{ $-3.3^{+1.3}_{-1.9}$  } \\
\hline
\multicolumn{5}{c}{Derived Parameters}\\
\hline
$a/R_{\star}$ &  $16.95^{+0.34}_{-0.82}$   &  $34.38^{+0.69}_{-2.0}$   &  $16.75^{+0.47}_{-0.74}$   &  $36.1^{+1.1}_{-1.7}$  \\
$i$ ($^{\circ}$) &  $89.38^{+0.43}_{-0.64}$   &  $89.147^{+0.069}_{-0.2}$   &  $89.1^{+0.59}_{-0.69}$   &  $89.51^{+0.34}_{-0.35}$  \\
$\delta$ (\%) &  $0.0467^{+0.0024}_{-0.0023}$   &  $0.0722^{+0.0038}_{-0.0037}$   &  $0.047^{+0.0029}_{-0.0025}$   &  $0.0696^{+0.0041}_{-0.0038}$  \\
$T_{14}$ (days) &  $0.134^{+0.0014}_{-0.0013}$   &  $0.1695^{+0.0015}_{-0.0013}$   &  $0.133^{+0.018}_{-0.02}$   &  $0.17 \pm 0.031$  \\
$T_{23}$ (days) &  $0.1279^{+0.0013}_{-0.0012}$   &  $0.1573^{+0.0014}_{-0.0015}$   &  $0.127^{+0.016}_{-0.02}$   &  $0.159 \pm 0.029$  \\
$g_{1,1}$ & \multicolumn{2}{c}{ $0.402^{+0.063}_{-0.06}$  } & \multicolumn{2}{c}{ $0.403^{+0.062}_{-0.059}$  } \\
$g_{2,1}$ & \multicolumn{2}{c}{ $0.143^{+0.058}_{-0.054}$  } & \multicolumn{2}{c}{ $0.139^{+0.058}_{-0.054}$  } \\
$g_{1,\mathrm{RM}}$ & \multicolumn{2}{c}{ $0.562^{+0.094}_{-0.093}$  } & \multicolumn{2}{c}{ $0.558^{+0.091}_{-0.094}$  } \\
$g_{2,\mathrm{RM}}$ & \multicolumn{2}{c}{ $0.162^{+0.088}_{-0.081}$  } & \multicolumn{2}{c}{ $0.161^{+0.087}_{-0.082}$  } \\
$e$ & 0 (fixed) & 0 (fixed) &  $0.085^{+0.179}_{-0.067}$   &  $0.114^{+0.204}_{-0.068}$  \\
$R_p$ ($R_\oplus$) & $ 2.15\pm 0.10$ & $ 2.67\pm 0.12$ & $ 2.15\pm 0.10$ & $ 2.64\pm 0.12$\\
$a$ (AU)$^a$ & $ 0.0719^{+ 0.0031}_{-0.0044}$ & $ 0.1458^{+ 0.0062}_{-0.0101}$ & $ 0.0710^{+ 0.0033}_{-0.0041}$ & $ 0.1531^{+ 0.0074}_{-0.0092}$\\
\enddata
\tablenotetext{a}{$R_p$ derived using the $R_*$ value from Table~\ref{tab:prop}}
\end{deluxetable*}

The GP fit accurately reproduced the overall out-of-transit variability (Figure \ref{fig:LCplot}). Similarly, the resulting period matched the rotation period from Section~\ref{sec:member} (6.54\,d versus 6.45\,d), and  agrees with the predicted value for the star's mass and membership to UMaG ($6.9\pm0.4$\,d). 

The transit duration suggests a low eccentricity for both planets ($e<0.2$), as is common for multitransiting systems \citep{Van-Eylen2015}. Our analysis did not include any correction for biases in the eccentricity distribution of planets \citep{2014MNRAS.444.2263K}, but accounting for this would only drive the resulting eccentricity values down. Consistent with this, the $\rho_*$ derived from the transit  assuming $e=0$ and a uniform density prior yields $\rho_*=1.331^{+0.056}_{-0.1}\rho_\odot$, in excellent agreement with our derived value from Section~\ref{sec:star} ($\rho_*=1.3\pm0.15\rho_\odot$). Thus, if we were to assume that the eccentricities are $\simeq$0, we can consider this an additional verification of our adopted stellar parameters.

Owing to the relatively low amplitude of the RM effect for \name~b and the complication of stellar activity, our posterior for the sky-projected spin-orbit misalignment $\lambda$ is broad. However, we clearly demonstrated that the planetary orbit is prograde; retrograde orbits would yield $|\lambda|>90^{\circ}$, which is completely ruled out (Figure~\ref{fig:transcorner}). We discuss the implications of this measurement in more detail in Section~\ref{sec:discussion}. Furthermore, we clearly detected the RM effect owing to the transit of \name~b (the signal is inconsistent with no RM signal), confirming that the planet is real. 

As an additional test on our RM fit, we ran an MCMC chain using a linear trend in velocity rather than a second-order polynomial (i.e., $\ddot{\gamma}$ fixed at 0). The resulting $\lambda$ posterior was not significantly different ($\lambda=7.0\pm35^\circ$). The second-order polynomial was preferred statistically ($\Delta BIC = 5$), so we use it for all results reported in Table~\ref{tab:transfit}.

\section{False-positive Analysis}\label{sec:fpp}
While planet b was confirmed through detection of the RM signal, we have no such detection for planet c. We instead validated the planet statistically by considering the three false-positive scenarios below. 

\subsection{Eclipsing Binary}
We compared the wealth of radial velocity data (Table~\ref{tab:RVs}) to the predicted velocity curve of a planet or binary at the orbital period of the outer planet (20.5\,d). We assumed a low eccentricity ($e<0.1$) and sampled over the whole range of $\omega$ and mass ratios. We included a zero-point correction term between instruments, which takes the value preferred by the predicted/model velocity curve. Including this term meant that 2-3 epochs each from TRES and NRES provided little information, and were not included. To account for stellar jitter and instrumental drift, we inflated errors in the velocities based on the scatter between points for each instrument (37\,\mps\ for SOPHIE/ELODIE and 24\,\mps\ for Lick). 

\begin{figure}[htp]
    \centering
    \includegraphics[width=0.47\textwidth]{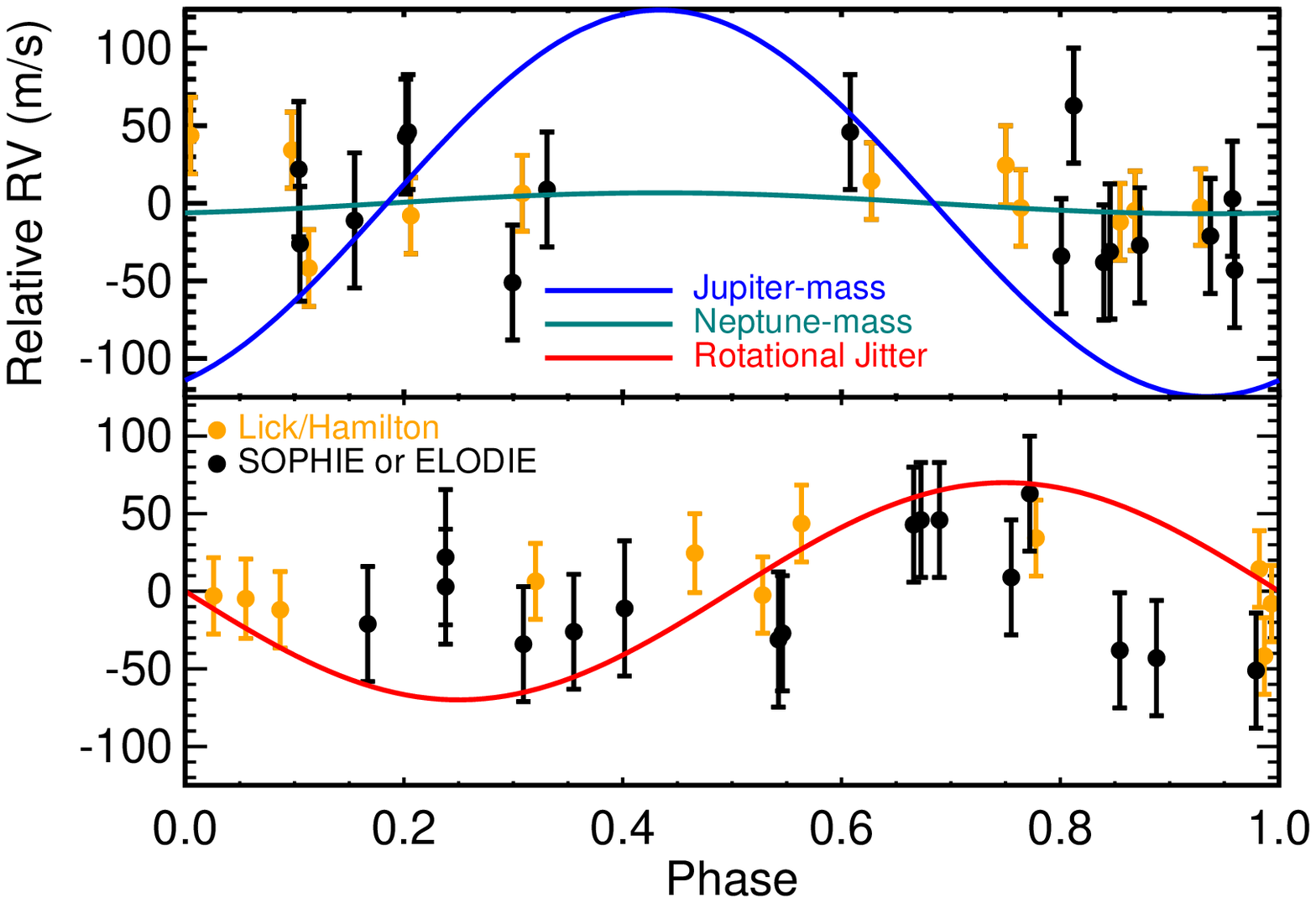}
    \caption{Top: velocities from Lick/Hamilton (orange) and SOPHIE or ELODIE (black) compared to the expected curve for a Jupiter- (blue) or Neptune-mass (teal) planet matching the orbital period of planet c. Bottom: the same velocities folded to the rotation period of the star and the approximate rotational jitter (red) expected from the {\it TESS} light curve and \vsini. Measurement have been inflated from their reported values based on the scatter between points. }
    \label{fig:rvs}
\end{figure}

The velocities are not precise enough to detect either planet, but easily rules out (at 99.7\%) any stellar or planetary companion down to $\simeq$Jupiter-mass at the orbital period of planet c (Figure~\ref{fig:rvs}).

\subsection{Background Eclipsing Binary}
 As detailed in \citet{Vanderburg2019}, if the observed transits are due to blends from a background eclipsing/transiting system, the true radius ratio can be determined from the ratio of the ingress time ($T_{12}$ or $T_{34}$) to the time between first and third contact ($T_{13}$). This provides a constraint on the brightest possible background source that could produce the observed transit depth: $\Delta m_{\rm{TESS}} \leq 2.5\log_{10}(T^2_{12}/T^2_{13}/\delta)$, where $\delta$ is the transit depth. Using our results from Section \ref{sec:transit}, we find $\Delta m < 2.4$\,mag at 99.7\% confidence. The combination of AO imaging and \gaia\ DR2 (Section~\ref{sec:AO}) rule out any such background star down to $<$80\,mas and out to 4.5\arcmin. 
 
 We also rule out a background eclipsing binary behind \name. The high-resolution imaging spans 9 years (1999.16--2008.28) and is sensitive to companions brighter than our magnitude threshold down to 80\,mas. Due to its proper motion, \name\ has moved more than 100\,mas over the same time period; thus, any foreground or background star not visible in the earliest dataset would be visible in the final observation. 

\subsection{Companion Eclipsing Binary}
To explore the range of possible stellar companions, we used a Monte Carlo simulation of 5$\times10^6$ binaries, comparing each generated system to the velocities, high-resolution imaging, and limits from \gaia\ imaging and astrometry. Companions were generated following a log-normal distribution in period following \citep{Raghavan2010}, but uniform in other orbital parameters. The radial velocity data listed in Table~\ref{tab:RVs} span more than 22 years, which overlaps in parameter space with the high-contrast imaging data (down to $\simeq$1\,AU) and \gaia\ constraints. A negligible fraction ($<0.01\%$) of generated companions are consistent with the external constraints, resolved in \tess, and reproduce the observed transit depth, statistically ruling out this scenario.

\section{Dynamical Analysis}\label{sec:dyn}

Studying the dynamical state of exoplanetary systems provides insights into the interactions between planets, their orbital evolution, and the possibility of additional, undetected planets in the system \citep[e.g.,][]{li2014,kane2019e}. The latter of these is particularly important for compact planetary systems, which are commonly dynamically filled \citep{fang2013}. To investigate these effects, we used the Mercury Integrator Package \citep{chambers1999} to conduct N-body integrations of the system. For this analysis, we adopt the stellar properties provided by Table~\ref{tab:prop} and the planetary properties (for the $e$, $\omega$ float case) provided by Table~\ref{tab:transfit}. We used the methodology described by \citet{kane2015b,kane2019c}, which both explores the intrinsic dynamical stability of the system using the observed parameters, and inserts additional planets to test the viability of possible additional planets. The time step of the integrations was set to 0.2~days in order to adequately sample the orbital period of the inner planet \citep{duncan1998}. Since only the planetary radii are provided by the measurements described in this work, we estimate the planetary masses using the probabilistic forecasting method of \citet{chen2017}. For both planet b and c, these masses are computed as 5.5 and 7.3 $M_\oplus$ respectively. We then executed an initial single dynamical simulation for $10^7$ simulation years that demonstrated the observed orbital architecture is a stable configuration. The chosen time step of 0.2~days maintained an energy conservation error of $dE/E \sim 10^{-9}$. To test for further stable locations, we inserted an Earth-mass planet at locations within the range 0.05--0.18~AU, which encompasses the semi-major axes of both planets (see Table~\ref{tab:transfit}). This process sampled several hundred locations within that range with the simulated planet placed at random starting locations. The results of this suite of simulations revealed that there is a stable island where an additional Earth-mass planet could be harbored located in the semi-major axis range of 0.099-0.112~AU. Even if present and transiting, such a planet is below the detection sensitivity of the photometry (Figure~\ref{fig:inject}). To fully explore the range of allowable planetary masses between planets b and c, we repeated the above analysis with an inserted Neptune, Saturn, and Jupiter-mass planet. The stable island between the planets were largely retained for the Neptune and Saturn-mass injected planets, but completely disappeared for the Jupiter-mass case. Thus Jupiter-mass planets are dynamically ruled out between the b and c planets, and transits of a Saturn-mass (Jupiter-radius) planet are ruled out by the photometry, but could still be present in a non-transiting capacity.

\section{Summary and Discussion}\label{sec:discussion}

We presented the discovery, characterization, and confirmation/validation of two planets transiting the bright ($V=6.9$ mag; Figure~\ref{fig:uniqueness}) star \name, a Sun-like star ($M_*=0.99\pm0.03M_\odot$). Based on its kinematics, lithium abundance, and rotation, we confirmed \name\ to be a member of the 414\,Myr old Ursa Major Moving Group. In addition to membership, we updated the stellar properties of \name\ based on the SED, \gaia\ DR2 distance, and existing high-resolution spectroscopy.  Using the {\it TESS} light curve, we determined the two planets have radii of $2.15\pm0.10R_\oplus$ and $2.67\pm0.12R_\oplus$ and periods of 7.11\,d and 20.54\,d, respectively. We simultaneously fit the {\it TESS} light curve with the HARPS-N spectroscopy of the RM effect taken during a transit of the inner planet. In addition to confirming the planet, the HARPS-N data demonstrate that the planet has a prograde orbit. Lastly, we validated the outer planet by ruling out non-planetary explanations for the observed signal. 

The two planets around \name\ add to the growing number of known transiting planets around young stars that are members of young ($<$1\,Gyr) clusters or moving groups \citep{Rizzuto2017, Curtis2018}. As the sample grows, it will enable studies into the evolution of planetary systems through the statistical comparison of young and old planetary systems, which in turn yields information about how exoplanets evolve. 

\name~b is the second young small planet with a published measurement of its spin-orbit alignment, after DS Tuc~b \citep{Zhou20, Montet20}, and the first in a multiplanet system. Both DS Tuc~b and \name\,b show prograde orbits. However, with the data currently in hand, our constraints on the spin-orbit alignment are poor. Further RM observations of multiple transits to increase the overall signal-to-noise and average over the effects of stellar activity will allow us to measure $\lambda$ more precisely. Nonetheless, \name\ is consistent with the trend of aligned orbits for compact multiplanet systems \citep{Albrecht13, Zhou18}, with only a few exceptions \citep{Huber13,Dalal19}. 

We found in Section~\ref{sec:inc} that the stellar rotation axis is likely to lie approximately in the plane of the sky. Taken together with the sky-projected spin-orbit misalignment, this suggests that the planets are aligned in 3D. Indeed, using Eqn.~7 of \cite{Winn:2007} and our measured values of the stellar and planetary inclinations and the spin-orbit misalignment, we calculate a 3-dimensional spin-orbit misalignment of $\psi<50^{\circ}$ at $1\sigma$ confidence. A more precise future measurement of $\lambda$ would also allow better constraints on $\psi$, as this is currently the limiting factor on the precision of $\psi$.

\subsection{Prospects for Follow-up}

\begin{figure*}[htp]
    \centering
    \includegraphics[width=0.94\textwidth]{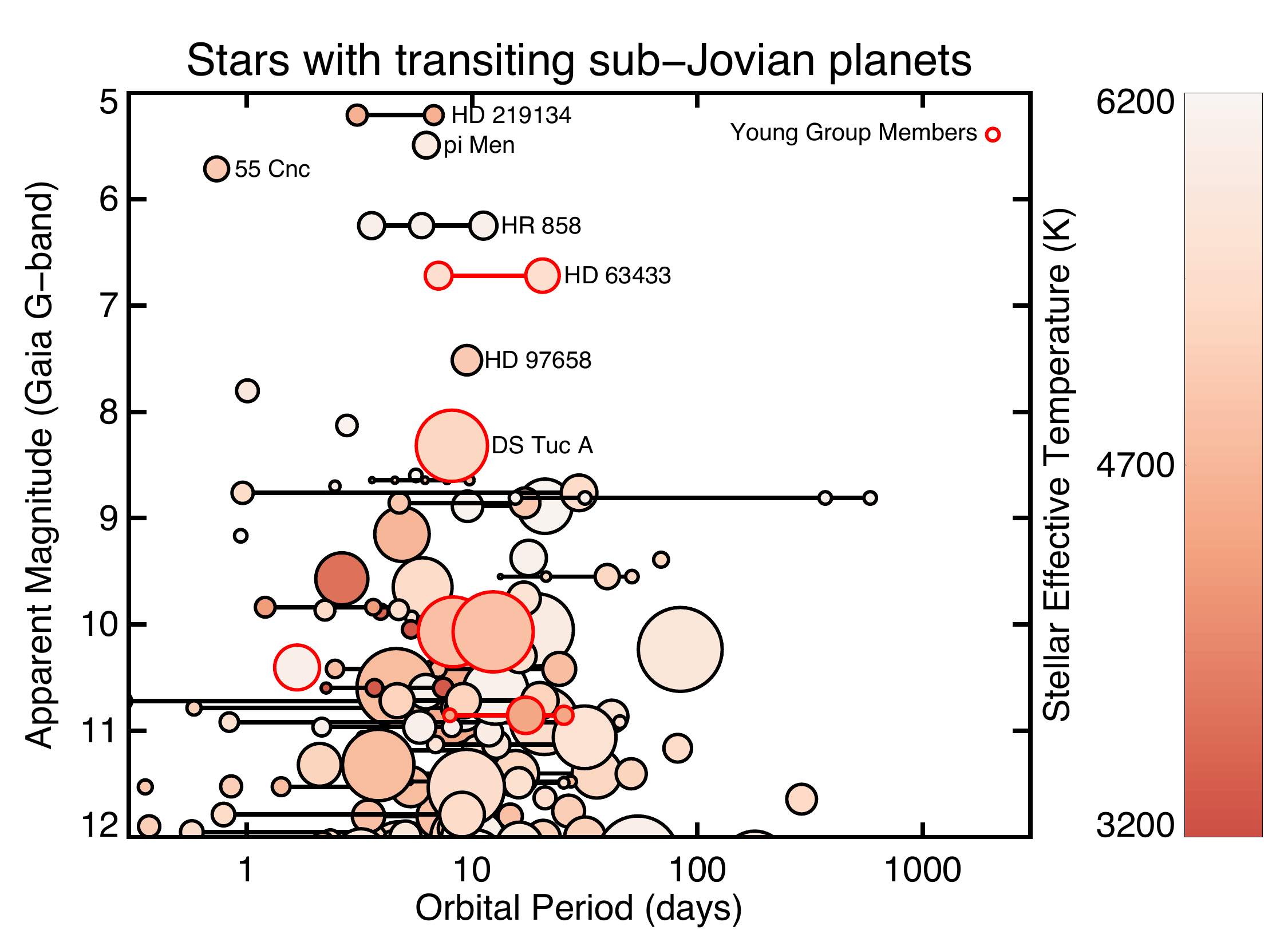}
    \caption{\name\ in context with the population of known small ($R_p<7$ \re) transiting planets, in terms of orbital period and host star brightness. Symbol size is proportional to the planetary radius, and symbol color to stellar effective temperature; planets transiting the same star are connected by lines, and non-transiting planets in these systems are not depicted. Several of the brightest systems are labeled; those in bold are \tess\ discoveries, while those in normal type were previous discoveries. We highlight planets in young ($<1$ Gyr) clusters and associations in red. \name\ is among the brightest stars known to host transiting planets, and is a prime target for a variety of follow-up observations. }
    \label{fig:uniqueness}
\end{figure*}

Thanks to \name's brightness ($V=6.9$, $K=5.3$; see Figure~\ref{fig:uniqueness}), this system is ideal for a variety of follow-up observations to characterize the planets and the system as a whole. Observations over the coming years will allow us to determine the system's 3-D architecture, measure mass loss from the planets and study their atmospheres, and potentially measure the masses of the planets.  

The HARPS-N observations of the b transit demonstrate that \name\ is well suited for additional RM observations. Repeat observations of the planet b would enable more detailed accounting of stellar variability \citep{Zhou20, Montet20} and provide more robust constraints on $\lambda$. Observations of the c transit would both confirm the planet and allow a measurement of the mutual inclination between the orbits of the two planets.

One of the first discoveries from the young planet population has been that young planets are statistically larger than their older counterparts \citep[e.g.,][]{Rizzuto2018,Mann:2018}. This offset could be explained by thermal contraction of an H/He dominated atmosphere \citep{2014ApJ...792....1L}, atmospheric mass-loss from interactions with the (still-active) host-star \citep{2009ApJ...693...23M}, or photochemical hazes making the atmosphere larger and puffier \citep{Gao_hazes}. Right now, the difference is only an offset in the planet radius distribution with age, making it difficult to distinguish between these scenarios. Instead, planet masses (and hence densities) are needed. While challenging, both of \name's planets may be within reach of existing PRV spectrographs. Assuming masses of $5.5M_\oplus$ and $7.3M_\oplus$ for planets b and c respectively \citep{chen2017}, the predicted radial-velocity amplitudes are $\simeq$2\,\mps. This signal is within the reach of existing instruments, but still much smaller than the estimated stellar jitter (20--30\,\mps; Figure~\ref{fig:rvs} and Table~\ref{tab:RVs}). The planet b is especially challenging given the similarity of its orbital period to the stellar rotation period (7.11\,d versus 6.45\,d). However, a focused campaign designed to separate planetary and stellar signals, as was done for the young system K2-100 \citep{2019MNRAS.490..698B}, will likely yield a mass constraint for planet c. 

\citet{Wang2019} and \cite{Gao_hazes} argue that young planets are likely to have flat transmission spectra due to either dust or photochemical hazes. There is some evidence to support this from transmission spectroscopy follow-up of young systems \citep{Libby-Roberts:2020, Thao2020}. However, a wider set of observations are required to explore under what conditions young atmospheres are dominated by hazes, dust, and/or clouds. Because the host is bright ($H\simeq5$), both planets are well within reach of transmission spectroscopy with {\it HST} or {\it JWST}.

Both of the planets lie on the large-radius, gas-rich side of the radius valley \citep[e.g.,][]{2013ApJ...775..105O, FultonPetigura18}. Given the young age of the system, it is likely that both planets are actively losing their atmospheres through photoevaporation \citep[e.g.][]{Owen2012, 2013ApJ...776....2L} or core-powered mass-loss \citep{GinzburgCorepowered2018}. Given the X-ray flux of \name\ observed by XMM-Newton as a part of its slew catalog ($94$ erg s$^{-1}$ cm$^2$) and the energy-limited mass-loss relation \citep[e.g.][]{OwenAtmospheric2019}, we estimate mass-loss rates of $\eta \approx 2.79\times 10^{11}$ g s$^{-1}$ and $\eta \approx 7.07\times 10^{10}$ g s$^{-1}$ for b and c, respectively, where $\eta$ describes the heating efficiency of the atmospheres. This is higher than many other planets of similar size, including Gl 436b and GJ 3470b, both of which have detected exospheres \citep{Ehrenreich2015, Ninan:2020}.

\acknowledgments
The authors thank the anonymous referee for their careful reading and thoughtful comments on the manuscript. The THYME collaboration would like to acknowledge Bandit, Erwin, Charlie, Edmund, Wally, and Marlowe for keeping us happy during the preparation of this paper. AWM would like to thank the COVID-19 response team at UNC Hospital in Chapel Hill. We thank Josh Schlieder for conversations on the properties of UMaG and the membership of \name. RG, TJ, DL, MK, MO, HS and IT gratefully acknowledge Allan R. Schmitt for making his light curve-examining software LcTools freely available.

AWM was supported through NASA’s Astrophysics Data Analysis Program (80NSSC19K0583). MLW was supported by a grant through NASA's \ktwo\ GO program (80NSSC19K0097). This material is based upon work supported by the National Science Foundation Graduate Research Fellowship Program under Grant No. DGE-1650116 to PCT. AV's work was performed under contract with the California Institute of Technology/Jet Propulsion Laboratory funded by NASA through the Sagan Fellowship Program executed by the NASA Exoplanet Science Institute. D. D. acknowledges support from NASA through Caltech/JPL grant RSA-1006130 and through the TESS Guest Investigator Program Grant 80NSSC19K1727.

This paper includes data collected by the TESS mission, which are publicly available from the Mikulski Archive for Space Telescopes (MAST). Funding for the TESS mission is provided by NASA’s Science Mission directorate. This research has made use of the Exoplanet Follow-up Observation Program website, which is operated by the California Institute of Technology, under contract with the National Aeronautics and Space Administration under the Exoplanet Exploration Program. This work has made use of data from the European Space Agency (ESA) mission \emph{Gaia} \footnote{\url{https://www.cosmos.esa.int/gaia}}, processed by the \emph{Gaia} Data Processing and Analysis Consortium (DPAC)\footnote{\url{https://www.cosmos.esa.int/web/gaia/dpac/consortium}}. Funding for the DPAC has been provided by national institutions, in particular the institutions participating in the \emph{Gaia} Multilateral Agreement.  This research has made use of the VizieR catalogue access tool, CDS, Strasbourg, France. The original description of the VizieR service was published in A\&AS 143, 23. Resources supporting this work were provided by the NASA High-End Computing (HEC) Program through the NASA Advanced Supercomputing (NAS) Division at Ames Research Center for the production of the SPOC data products. We acknowledge the use of public TOI Release data from pipelines at the TESS Science Office and at the TESS Science Processing Operations Center. Based on data retrieved from the SOPHIE archive at Observatoire de Haute-Provence (OHP), available at atlas.obs-hp.fr/sophie. This work makes use of observations from the LCOGT network. Based on observations made with the Italian {\it Telescopio Nazionale Galileo} (TNG) operated by the {\it Fundaci\'on Galileo Galilei} (FGG) of the {\it Istituto Nazionale di Astrofisica} (INAF) at the {\it Observatorio del Roque de los Muchachos} (La Palma, Canary Islands, Spain). Part of this research was carried out at the Jet Propulsion Laboratory, California Institute of Technology, under a contract with the National Aeronautics and Space Administration (NASA). The HARPS-N project has been funded by the Prodex Program of the Swiss Space Office (SSO), the Harvard University Origins of Life Initiative (HUOLI), the Scottish Universities Physics Alliance (SUPA), the University of Geneva, the Smithsonian Astrophysical Observatory (SAO), and the Italian National Astrophysical Institute (INAF), the University of St Andrews, Queens University Belfast, and the University of Edinburgh.

\vspace{5mm}
\facilities{TESS, LCOGT 1m (Sinistro), LCOGT 1m (NRES), SMARTS 1.5m (CHIRON), Tillinghast 1.5m (TRES), SOAR (Goodman), TNG (HARPS-N), OHP 1.93m (ELODIE), OHP 1.93m (SOPHIE), Shane 3m (Hamilton), \ldots}

\software{\texttt{LcTools}, \texttt{misttborn.py}, \textit{emcee} \citep{Foreman-Mackey2013}, \textit{batman} \citep{Kreidberg2015}, matplotlib \citep{hunter2007matplotlib}, \texttt{corner.py} \citep{foreman2016corner}, \texttt{AstroImageJ} \citep{Collins17}, BANZAI \citep{McCully18}
          }

\clearpage

\bibliography{fullbiblio.bib}{}
\bibliographystyle{aasjournal}

\end{document}